\documentclass[useAMS,usenatbib]{mn2e}
\usepackage{graphicx}

\title[Dynamic Collapses of Spherical Black Holes]
{Relativistic Self-similar Dynamic Collapses of\\
 Black Holes in General Polytropic Spherical Clouds}
\author[Biao Lian and Yu-Qing Lou]{Biao
Lian$^{1,\ 2}$
 and Yu-Qing Lou$^{1}$
\thanks{louyq@mail.tsinghua.edu.cn}\\
$^{1}$Department of Physics and Tsinghua Centre for Astrophysics
(THCA), Tsinghua University, Beijing 100084, China\\
$^{2}$Department of Physics, McCullough Building,
Stanford University, Stanford, California 94305-4045, USA
 }
\begin{document}

\date{Accepted 2013 November 21. Received 2013 November 21; 
 in original form 2013 August 28}

\pagerange{\pageref{firstpage}--\pageref{lastpage}} \pubyear{2002}

\maketitle
\label{firstpage}

%\begin{abstract}
%We study the hydrodynamic self-similar accretion of polytropic gas spheres by black %holes. In order to approximately capture effects brought by general relativity (GR), %we adopt the Paczynski-Wiita potential as a more precise Newtonian approximation of %GR. A new parameter $s$ is introduced with a physical meaning of the square of the %ratio of sound speed and light speed. Solutions are found different for polytropic %exponent $\gamma>4/3$. Two (for small $s$) or no (for large $s$) expansion-wave %collapse solutions exist when $\gamma>4/3$, which represents the collapse of static %singular spheres. Spherical accretion is shown to be highly efficient that it may be %responsible for the quick formation of super-massive black holes in the early %universe, which may involve into hard $X-$ray sources or quasars. Self-similar %solutions are also used on the black hole formation during supernova explosion of a %massive star, the time scale of which is estimated around $10^{-3}s$. The rebound %shock in supernova is also discussed based on shock solutions.
%\end{abstract}
\begin{abstract}
We study the hydrodynamic self-similar mass collapses
 of general polytropic (GP) spherical clouds to central
 Schwarzschild black holes and void evolution with or
 without shocks.
In order to grossly capture characteristic effects of
 general relativity (GR) outside yet close to the
 event horizon of a
 Schwarzschild black hole and to avoid mathematical
 complexity, we adopt the approximation of the
 Paczynski-Wiita gravity to replace the simple
 Newtonian gravity in our model formulation.
A new dimensionless parameter $s$ appears with the
 physical meaning of the square of the ratio of the
 sound speed to the speed of light $c$.
Various self-similar dynamic solutions are constructed
 for a polytropic index $\gamma>4/3$.
Two (for small enough $s<1$) or no (for large enough $s<1$)
 expansion-wave collapse solutions (EWCSs) with central event
 horizons exist when $\gamma>4/3$, representing the collapse
 of static singular GP spheres towards
 the central singularity of spacetime.
Such GP spherical dynamic mass collapse is
 shown to be highly efficient for the rapid formation
 of supermassive black holes (SMBHs; mass range of
 $\sim 10^6-10^{10}M_{\odot}$) in the early universe
 or even hypermassive black holes (HMBHs; mass range of $\sim 10^{10}-10^{12}M_{\odot}$) if extremely massive mass
 reservoirs could be sustained for a sufficiently long
 time, which may evolve into hard $X-$ray/gamma ray
 sources or quasars according to their surroundings.
Self-similar dynamic solutions of a GP gas
 are also proposed for the stellar mass black hole formation
 during the violent supernova explosion of a massive
 progenitor star, the timescale
 of which is estimated of $\sim 10^{-3}$ seconds.
Rebound shocks travelling in supernovae are also discussed
 based on our self-similar shock expansion solutions.
\end{abstract}

%\begin{keywords}
%hydrodynamics---black hole physics---accretion, accretion discs---early %Universe---quasars: general---supernovae: general
%\end{keywords}

\section{Introduction}

Mass accretions and collapses of black holes, especially
 supermassive black holes (SMBHs), are commonly thought to be related to galaxy formations and powerful quasars, and are still not well understood.
To date, extensive observations have shown strong evidences
 that SMBHs of $10^6\sim 10^{10}M_{\odot}$
%  $10^6\sim 10^9$ solar masses
 reside in the corresponding galactic nuclei and manifest
 novel nucleus properties (e.g. Begelman et al. 1984; Rees 1984;
 Kormendy \& Richstone 1995; Magorrian et al. 1998; Marconi \& Hunt
 2003; Wandel 1999, 2002; Ferrarese \& Ford 2005; Graham \& Driver
 2007; Lou \& Jiang 2008; Lou \& Wu 2012; Shen et. al. 2009;
 McConnell et al. 2011; Shen et. al. 2011).
In particular, McConnell et al. (2011) indicate that the two
 ten-billion-solar-mass black holes at the centers of two giant
 elliptical galaxies NGC 3842 and NGC 4889 (typical brightest
 cluster galaxies -- BCGs) tend to be much more massive and
 deviate from the well-known $M_{\rm BH}-\sigma$ relation
 significantly (e.g. Lou \& Jiang 2008).
It is widely believed that SMBHs in galactic nuclei are
 closely related to the formation and violently active
 phenomina (AGNs and quasars) of their host bulges.
Recent observations of luminous quasars with red shifts as high
 as $z=7$ (e.g. Mortlock et. al. 2011) indicate SMBHs with
 $\sim 10^8M_{\odot}$ or higher must have occurred in one
 billion years or so after the Big Bang of the Universe.
% {\bf Evidence for $10^{10}M_{\odot}$?}
Another interesting recent development is the possible formation
 of intermediate mass black holes (IMBHs) at the centers of
 globular clusters and the tentative observational evidence
 for this plausible scenario (Lou \& Wu 2012 and references therein).
In the past several decades, many research works focusing on
 mass accretions towards black holes have been done, most of which
 were based on an accretion disk structure (e.g. Lynden-Bell 1969;
 Shakura \& Sunyaev 1973; Novikov \& Thorne 1973; Lynden-Bell \&
 Pringle 1974; Paczynski \& Wiita 1980; Rybicki \& Lightman 1979;
 Sunyaev \& Titarchuk 1980; Haardt \& Maraschi 1991, 1993; Shapiro
 et al. 1976).
We here explore an alternative yet complementary scenario.
In the very early universe before stages of galaxy
 formation, the distribution of cosmic matters may be much
 denser and structure-less.
Angular momentum may not be enough to induce disks in the
 formation of proto-galaxies.
The black hole growth process around such early stage may not
 be able to rely on a surrounding well-established accretion disk.
Loosely bound systems of sub-galactic arising from weakly
 interacting dark matter may have provided hosts for early SMBHs
 and quasars (e.g. Peebles 1982; Blumenthal et. al. 1984).
Inhomogeneity of the cosmological space time may also be present
 (e.g. Tolman 1934; Guth 1981; Mukhanova et. al. 1992; Lyth \&
 Riotto 1999), so that large-scale fluctuations and remaining
 acoustic waves from the Big Bang can in principle induce
 considerable local mass density fluctuations
 (e.g. Lukash 1980; Ford \& Parker
 1977), yielding high mass density peaks in the vast space.
Dense grossly spherical blobs with or without central
 black holes born in such
 environments might experience quite novel dynamic mass
 collapse processes (Hoyle \& Fowler 1963; Fowler 1964;
 Begelman 2009).
In addition to disk-type accretion models, grossly spherical
 symmetric
 dynamic mass collapses may become another significant
 yet simplified model for most such early
 black hole evolution processes,
 which is the focus of this paper.
In the literature, spherical accretion models like Bondi accretion
 are also used as simple estimation of mass accretion rates around
 compact objects (Bondi 1952; Bondi \& Hoyle 1944; Mocanu 2012).
We expect our dynamic spherical models would be useful for describing
 characteristics of mass inflows before the formation of a
 stable accretion disk structure inside the galactic nuclei.

There has been extensive investigations on the mass accretion
 and collapse of spherical
 gas clouds towards central massive objects through construction of
 spherical hydrodynamic self-similar solutions (e.g., Shu 1977;
 Whitworth \& Summers 1985; Lou \& Shen 2004; Lou \& Cao 2008;
 Wang \& Lou 2008).
Most researchers so far work in the framework of Newtonian gravity which
 cannot give an exact description around highly general relativistic
 compact objects such as black holes to allow for a massive compact
 object occupying a very much smaller spatial volume, sitting at the
 center of $r=0$.
The physical behaviours of dynamic mass accretions and collapses
 close to a black hole then cannot be described more precisely.
The construction of self-similar hydrodynamic gravitational
 collapse solutions using general relativity was done several
 years ago by Cai \& Shu (2005), which describes a collapse of
 relativistic singular isothermal sphere (SIS) which is
 initially unstable (Cai \& Shu 2003).
Such solution construction involving general relativity
 is complicated and is done for isothermal sphere with
 a polytropic exponent $\gamma=1$.
%However, since these works are based on Newtonian gravitation,
%the center massive objects in the solutions are usually
%infinitely small in volume, concentrating on the point $r=0$.
In order to remedy the deficiency of Newtonian gravity
 while at the same time circumvent the mathematical machinery
 of general relativity and study the gravitational collapse
 of GP spheres, we invoke the Paczynski-Wiita
 potential (Paczynski \& Wiita 1980) as a first-order correction
 of Newtonian gravity which is the weak field approximation
 of general relativity, and construct spherical self-similar
 gravitational collapse solutions of GP gas spheres.
Under the approximation of the Paczynski-Witta potential, the
 compact object has a finite Schwarzschild radius, and many
 properties of spacetime near a black hole are reproduced
 approximately (Paczynski \& Witta 1980; Abramowicz 2009).

%When extremely relativistic conditions are imposed, we find solutions governed by Paczynski-Wiita potential significantly distinct from those in the framework of Newtonian gravitation. Such conditions may occur in a supernova collapse or explosion process in accompany with the formation of a nascent neutron star or black hole (Chandrasekhar 1935; Oppenheimer \& Volkoff 1939). Highly relativistic process will involve tempestuous changes in space time metric as well as gas inflows and need to be described with the whole participation of general relativity (Oppenheimer \& Snyder 1939; Tolman 1939; Misner \& Sharp 1964,1965; Misner 1965). A realistic supernova processes involve many more physics including radiation, heat transfer and production or annihilation of leptons and neutrinos (Bethe 1990), and thus is still a big unsolved problem. Despite of so many complexities in supernova physics, we wish to focus on the hydrodynamics part, and catch some basic features of the supernova such as the time scale of the formation of the stellar black hole.

When relativistic conditions are introduced, solutions
 governed by the Paczynski-Wiita potential are significantly
 distinct from those in the framework of Newtonian gravity.
Such conditions may occur in stellar core collapses and
 supernova explosion processes in accompany with the
 formation of a nascent neutron star or black hole (e.g.
 Chandrasekhar 1935; Oppenheimer \& Volkoff 1939).
Such highly relativistic dynamic processes will involve
 tempestuous changes in spacetime metric as well as gas
 inflows and need to be described by general relativity
 (Oppenheimer \& Snyder 1939; Tolman 1939;
 Misner \& Sharp 1964, 1965; Misner 1965).
Realistic supernova processes also involve many other physics
 including radiation, heat transfer and production or annihilation
 of leptons and neutrinos (e.g. Bethe 1990), and thus remains
 still a major unsolved astrophysical problem.
%Despite of so many complexities in supernova physics, we wish to
% give some na\"ive estimates of such processes such as the time
% scale of the formation of the stellar black hole through our
% self-similar hydrodynamic formulation.
Despite of much complexities in supernova physics, we wish
 to focus on the hydrodynamic part, give several estimates,
 and catch some basic features of the supernova such as the
 timescale of the formation of a stellar black hole.

%The rest of the paper is organized as the following: In section 2 the basic equations and reduced ODEs for constructing self-similar solutions are introduced and derived. In section 3 we consider some asymptotic solutions of the ODEs, seizing the key differences of solutions in our work compared to those obtained using Newtonian gravitations. Explicit numerical solutions of various kinds including shock waves are shown and analyzed in section 4. The physical consequences closely related to the self-similar solutions are analysed in section 5. Finally, the conclusion is summarized in section 6.

The remainder of the paper is structured as follows.
 In Section 2, the basic nonlinear hydrodynamic partial
 differential equations (PDEs) and the corresponding
 reduced ordinary differential equations (ODEs) for
 constructing the important class of self-similar
 hydrodynamic solutions are introduced and derived.
In Section 3, we consider several asymptotic analytic
 solutions of the coupled nonlinear ODEs, indicating
 key differences of solutions in our model formalism
 as compared to those obtained by others previously
 using the Newtonian gravity.
Explicit numerical solutions of various kinds including
 shock waves and/or central voids are shown and
 analyzed in Section 4.
Section 5 describes the physical consequences closely related
 to self-similar hydrodynamic solutions obtained in this
 paper, and conclusions are summarized in Section 6.
Three appendices of more detailed mathematical analysis
 are attached for the convenience of references.

%\section{Basic equations of the model}
\section{The Basic Model Description}

%In our model the force of Paczynski-Wiita potential is adopted instead of the Newtonian gravitational force to include the correction from general relativity. the Paczynski-Wiita force per unit mass takes the following form:
%\begin{equation}\label{force}
%g=-\frac{GM}{(r-2GM/c^2)^2}\ .
%\end{equation}
%where $G=6.67\times10^{-11}kg^{-1}\cdot m^3\cdot s^{-2}$ is the universal gravitation constant, $r$ is the radius from the center, $c$ is the speed of light and $M$ is the total mass enclosed inside the sphere with radius $r$. The term $2GM/c^2$ is the Schwarzschild radius of a Schwarzschild black hole with mass $M$. One should note the force $g$ here does not come from the gradient of the potential $\phi=-GM/(r-2GM/c^2)$ as is written by Paczynski and Wiita(1980), since $M$ is now a function of the radius $r$. One also note that it is an approximation that the force $g$ only depends on the mass $M$ enclosed in radius $r$, since the Poisson equation in the Newtonian theory breaks down. In order that the approximation is valid, the density of a hydrodynamic solution is required to be sufficiently low (decaying no slower than $r^{-2}$, see section \ref{thermal}) at $r\rightarrow\infty$.

In our model formalism, the gravity force under the
 approximation of the Paczynski-Wiita potential is
 adopted instead of the Newtonian gravity force
%satisfying the Poisson equation in order
 to include certain effects of general relativity in a
 simplified manner.
In our formalism, the Paczynski-Wiita gravity force per unit mass
 is
\begin{equation}\label{force}
g=-\frac{GM}{(r-2GM/c^2)^2}\ ,
\end{equation}
where $G=6.67\times10^{-8}$g$^{-1}$ cm$^3$ s$^{-2}$ is
 the universal gravitational constant, $r$ is the radius
 from the center, $c=3\times 10^{10}\hbox{ cm s}^{-1}$ is
 the speed of light and $M$ is the total mass enclosed
 inside a mass sphere of radius $r$.
In reference to the Newtonian gravity, the extra term
 $2GM/c^2$ included here is the Schwarzschild radius
 for a non-rotating Schwarzschild black hole of an
 enclosed mass $M$.
%{\bf
We emphasize that the specific gravity force $g$ here is
 simply the gradient of potential $\phi=-GM/(r-2GM/c^2)$ of
 Paczynski \& Wiita (1980) for a constant enclosed mass $M$.
As this expression of Paczynski-Wiita potential $\phi$
 is not consistent with the self-gravity as stipulated
 by the Poisson equation relating the gravitational
 potential $\phi$ and the mass density $\rho$,
%This force $g$ is actually still not quite exact,} since we have
 we no longer require the Poisson equation which
 ensures that the gravity force is produced
 solely by the enclosed mass $M$.
To justify our approximation, we demand the mass
 density being sufficiently low (at most
 proportional to $r^{-2}$, see subsection \ref{thermal})
 in the asymptotic regime of $r\rightarrow +\infty$.

%{\bf
We emphasize that the Paczynski-Wiita potential captures
  well the deviation from the $r^{-1}$ law of the Newtonian gravitational potential for an extremely dense object
  (e.g. Abramowicz 2009) -- an essential element of our
  paper, and thus can reproduce the characteristic
  features of general relativity.
Combined with non-relativistic particle kinetics in
  circular Keplarian orbits, it gives exactly the
  same radii of the marginally stable orbit
  ($r_{\rm ms}$) and the marginally bound orbit
  ($r_{\rm mb}$) as those given by Einstein's
  general relativity (Paczynski \& Wiita 1980),
  which justifies the validity of this ``pseudo
  Newtonian" approximation.
For the historical background, theoretical
  development and a step-by-step ``derivation" of the
  Paczy\'nski-Wiita potential as well as comparisons
  with Einstein's general relativity, the interested
  reader is referred to the two-page commentary of
  Abramowicz (2009) for more details.
There are other proposed potentials (e.g., Nowak \&
  Wagoner 1991; Semer\'ak \& Karas 1999;
  Klu\'zniak \& Lee 2002; Stuchl\'ik \& Kov\'ar 2008)
  with respective merits to capture features of general relativity, we adopt the Paczy\'nski-Wiita potential
  here mainly for beauty and simplicity.
% }

%In spherical coordinates $(r,\theta,\phi)$, the basic equations for a spherical evolution of a general polytropic gas are as follows
%\begin{equation}\label{basic1}
%\frac{\partial \rho}{\partial t}+\frac{1}{r^2}\frac{\partial}{\partial r}(r^2\rho u)=0\ ,
%\end{equation}
%\begin{equation}\label{basic2}
%\frac{\partial M}{\partial t}+u\frac{\partial M}{\partial r}=0\ ,
%\end{equation}
%\begin{equation}\label{basic3}
%\frac{\partial M}{\partial r}=4\pi r^2\rho\ .
%\end{equation}
%\begin{equation}\label{basic4}
%\frac{\partial u}{\partial t}+u\frac{\partial u}{\partial r}=-\frac{1}{\rho}\frac{\partial p}{\partial r}+g\ ,
%\end{equation}
%where $t$ is the time, $\rho\ ,p\ , u$ are the gas density, gas pressure and the radial flow velocity of the gas, respectively, and the force per %unit mass $g$ is given above in equation (\ref{force}). By further demanding the conservation of specific entropy, we have to add another equation:
%\begin{equation}\label{basic5}
%\Big(\frac{\partial}{\partial t}+u\frac{\partial}{\partial r}\Big)\ln\frac{p}{\rho^\gamma}=0\ ,
%\end{equation}
%where $\gamma$ is the polytropic index of the gas.

In a reference framework of spherical polar
 coordinates $(r,\ \theta,\ \phi)$, the basic
 nonlinear hydrodynamic PDEs for a spherically
 symmetric dynamic evolution of a
 GP gas are given below in the forms of
\begin{equation}\label{basic1}
\frac{\partial \rho}{\partial t}+\frac{1}{r^2}\frac{\partial}
{\partial r}(r^2\rho u)=0\ ,
\end{equation}
\begin{equation}\label{basic2}
\frac{\partial M}{\partial t}+u\frac{\partial M}{\partial r}=0\ ,
\end{equation}
\begin{equation}\label{basic3}
\frac{\partial M}{\partial r}=4\pi r^2\rho\ ,
\end{equation}
\begin{equation}\label{basic4}
\frac{\partial u}{\partial t}+u\frac{\partial u}{\partial r}
 =-\frac{1}{\rho}\frac{\partial p}{\partial r}+g\ ,
\end{equation}
where $t$ is the independent time variable, $\rho$, $p$ and $u$
 are the gas mass density, the gas pressure and the radial flow
 velocity of the gas, respectively, and the gravity force per
 unit mass $g$ is given by equation (\ref{force}) in
 the Pazcynski-Wiita approximation.
By further demanding the conservation of specific entropy
 along streamlines, we have
% to add another equation:
\begin{equation}\label{basic5}
\bigg(\frac{\partial}{\partial t}+u\frac{\partial}
 {\partial r}\bigg)\ln\bigg(\frac{p}{\rho^\gamma}\bigg)=0\ ,
\end{equation}
where $\gamma$ is the polytropic index of a dynamic gas.
%We note that
PDEs (\ref{basic2}) and (\ref{basic5})
 bear the same form which implies the involvement of
 enclosed mass in the GP EoS as discussed presently.

We now introduce in a self-consistent manner the following
 self-similarity transformation,
\[r=k^{1/2}t\ x\ ,\qquad\quad u=k^{1/2}v(x)\ ,\qquad\quad
 \rho=\frac{\alpha(x)}{4\pi Gt^2}\ ,\]
\begin{equation}\label{self-similar}
p=\frac{k\ \beta(x)}{4\pi Gt^2}\ ,\ \qquad\qquad
 M=\frac{k^{3/2}t\ m(x)}{G}\ ,
\end{equation}
such that nonlinear hydrodynamic PDEs
 (\ref{force})$-$(\ref{basic4}) can be
 consistently and readily
 reduced to a set of coupled nonlinear
% ordinary differential equations
 ODEs in terms of the independent
 self-similar variable $x$.
Here,
%$x$ is the independent self-similar variable, and
 dimensionless dependent variables $v(x),\
 \alpha(x),\ \beta(x),\ m(x)$ are four reduced
 functions of $x$ only.
The sound parameter $k$ is an allowed scaling
 factor to solve the coupled nonlinear ODEs.
The self-similar scaling index $n$ is simply
 set to unity from the very begining for a
 successful transformation
 (e.g. Wang \& Lou 2008; Lou \& Shi 2013).
For the mass conservation, nonlinear PDEs
 (\ref{basic2}) and (\ref{basic3}) together lead to
\begin{equation}\label{m}
m=(x-v)x^2\alpha\ ,
\end{equation}
and in order that the enclosed mass $M$ remains positive,
 the inequality $x-v>0$ must be met as dictated by physics.
For the point keeping $x-v=0$, the enclosed mass is zero and
 we would have an expanding central void.
The form of reduced pressure $\beta(x)$ is
 determined by PDE (\ref{basic5}) to be
\begin{equation}\label{beta1}
\beta={\cal C}\alpha^{\gamma}m^{2(\gamma-1)}\ ,
\end{equation}
where ${\cal C}$ is a constant proportional coefficient.
Here, we shall not consider the special case of
 $\gamma=4/3$ for a relativistically hot or
 degenerate gas (e.g., Lou \& Cao 2008);
 by properly adjusting the scaling factor $k$, we can always
 make the coefficient ${\cal C}=1$ without loss of generality,
 leading to the following simple algebraic form of $\beta(x)$
\begin{equation}\label{beta}
\beta=\alpha^{\gamma}m^{2\gamma-2}=\alpha^{3\gamma-2}
 (x-v)^{2\gamma-2}x^{4\gamma-4}\
\end{equation}
in reference to two algebraic relations
 (\ref{m}) and (\ref{beta1}).
The remaining two nonlinear PDEs (\ref{basic1})
 and (\ref{basic4}) give the following two coupled
 nonlinear ODEs
\begin{equation}\label{b1}
(x-v)\Big(\frac{2\alpha}{x}+\frac{\mbox{d}\alpha}
 {\mbox{d}x}\Big)=\alpha\frac{\mbox{d} v}{\mbox{d} x}\ ,
\end{equation}
\begin{equation}\label{b2}
\frac{1}{\alpha}\frac{\mbox{d}\beta}{\mbox{d} x}-(x-v)
 \frac{\mbox{d} v}{\mbox{d} x}=-\frac{(x-v)x^2\alpha}
 {[x-s(x-v)x^2\alpha]^2}\ ,
\end{equation}
where reduced pressure $\beta(x)$ is given
 by algebraic equation (\ref{beta}), and
 $s=2k/c^2$ is a constant parameter proportional to the square of
 the ratio of sound speed over the speed of light $c$.
Nonlinear ODEs (\ref{b1}) and (\ref{b2}) can be readily rearranged
% by a proper combination to adopt
 into the following form of
\[\Big[(x-v)^2-\gamma\frac{\beta}{\alpha}\Big]
 \frac{1}{\alpha}\frac{\mbox{d}\alpha}{\mbox{d}x}
 =\frac{(x-v)x^2\alpha}{[x-s(x-v)x^2\alpha]^2}\]
\begin{equation}\label{ode1}
\qquad\qquad\qquad
+\frac{2\gamma-2}{(x-v)}\frac{\beta}{\alpha}-\frac{2(x-v)^2}{x}\ ,
\end{equation}
\[\Big[(x-v)^2-\gamma\frac{\beta}{\alpha}\Big]\frac{1}{(x-v)}
 \frac{\mbox{d}v}{\mbox{d}x}=\frac{(x-v)x^2\alpha}
 {[x-s(x-v)x^2\alpha]^2}\]
\begin{equation}\label{ode2}
\qquad\qquad\qquad +\Big(\frac{2\gamma-2}{x-v}
 -\frac{2\gamma}{x}\Big)\frac{\beta}{\alpha}\ .
\end{equation}
Nonlinear ODEs (\ref{ode1}) and
 (\ref{ode2}) can be readily solved by standard
 numerical integration with specified asymptotic
 conditions and/or solutions which we elaborate next.

%{\bf Stop here!}

%\section{Asymptotic solutions}

%By some physical assumptions, asymptotic solutions inside certain regions can be obtained. These asymptotic behaviors are very helpful for understanding numerical solutions and finding particular solutions of interest.

\section{Asymptotic Analytic Solutions}

By some sensible physical assumptions and requirements,
 asymptotic analytical solutions within certain
 parameter regimes can be derived.
These asymptotic behaviors are very valuable and useful for
 constructing and understanding the numerical solutions.

\subsection{Asymptotic analytic solutions of finite\\
\hskip 0.9cm mass density and velocity at large $x$}

%{\bf Please check and compare with earlier known results!
% If necessary, please derive higher-order terms.
% August 24, 2013. Done Aug 27, 2013 }
For asymptotic analytic solutions with $\alpha(x)$
 and $v(x)$ approaching finite values at very
 large $x$, nonlinear ODE (\ref{b1}) in the
 leading order can be approximated as
\begin{equation}
\frac{2\alpha}{x}+\frac{\mbox{d}\alpha}{\mbox{d}x}=0\ ,
\end{equation}
yielding the asymptotic solution for
 $\alpha(x)$ at $x\rightarrow +\infty$
\begin{equation}\label{al}
\alpha=\frac{\alpha_0}{x^2}+o\Big(\frac{1}{x^2}\Big)\ .
\end{equation}
Asymptotic solution (\ref{al}) further leads to the leading
 order approximation of ODE (\ref{ode2}) at very large $x$ as
\begin{equation}\label{alf}
\frac{\mbox{d}v}{\mbox{d}x}=-\bigg[2\alpha_0^{3\gamma-3}
 -\frac{\alpha_0}{(1-s\alpha_0)^2}\bigg]\frac{1}{x^2}\ ,
\end{equation}
and the corresponding asymptotic solution
 of $v(x)$ at very large $x$ is therefore
\begin{equation}\label{18}
v(x)=v_0+\bigg[2\alpha_0^{3\gamma-3}-\frac{\alpha_0}
 {(1-s\alpha_0)^2}\bigg]\frac{1}{x}+o\Big(\frac{1}{x}\Big)\ .
\end{equation}
%{\bf Consistent with Lou \& Shi (2013 submitted); Aug 24, 2013}
For a realistic astrophysical consideration, we shall
 demand $s\alpha_0<1$ for the space-time domain outside
 the event horizon of a central Schwarzschild black hole.
This constraint is in fact due to the invalidation of the
 Paczynski-Wiita potential at high mass densities of infinite
 radius
% {\bf Need a clarification still! commented out Aug 25, 2013.}
 (the Poisson
 equation is no longer imposed here, see Section 2).
% {\bf Please clarify!}
In most realistic astrophysical flow systems, indeed, the gas
 density far away enough will always decrease sufficiently fast
 and go below this critical value of unity for $s\alpha_0$.
% {\bf What is this critical value?}
Asymptotic solution (\ref{18}) can be used to further obtain the
 expression of $\alpha(x)$ to the second order in the form of
\begin{equation}
\alpha=\frac{\alpha_0}{x^2}\Big\{1+\Big[2\alpha_0^{3\gamma-3}
 -\frac{\alpha_0}{(1-s\alpha_0)^2}\Big]\frac{1}{2x^2}\Big\}
 +o\Big(\frac{1}{x^4}\Big)\ .
\end{equation}
%Especially when $v_0=0$, and $\alpha_0$ meets the condition
%\begin{equation}\label{alpha0}
%2\alpha_0^{3\gamma-3}-\frac{\alpha_0}{(1-s\alpha_0)^2}=0\ ,
%\end{equation}
%we get an exact solution which describe
%the singular static polytropic sphere
%\[v=0\ ,\]
%\begin{equation}\label{singular}
%\alpha=\frac{\alpha_0}{x^2}\ ,
%\end{equation}
%Which is generally shown to be unstable in both the
%non-relativistic limit (Shu 1977; Lou \& Shen 2004)
%and relativistic case (Cai \& Shu 2003). It should
%be noted that the number of roots of equation
%(\ref{alpha0}) for $\alpha_0$ can be 0,1 or 2 (Figure
%\ref{fig:pcri}). For $\gamma<4/3$, it is shown that
%there is always only one root. When $\gamma>4/3$ and
%$s<(1-\sqrt{2}/2)=0.2929$, there is always two roots.
%When $s>0.2929$, there may be no root or two roots
%depending on the value of $\gamma(>4/3)$. This is
%important for understanding the properties of the
%critical lines of ODEs (\ref{ode1}) and (\ref{ode2}),
%and we will solve this problem when discussing the
%critical lines in section \ref{sect-cri}.
% {\bf Comparison with earlier results. Aug 25, 2013.}
%{\bf
By setting $s=0$ in the above two expressions, we
 readily recover the Newtonian results, and the
 above asymptotic analytical solutions carry
 the same form as those deduced in previous works
 with $n=1$ for the Newtonian gravity (e.g. Wang \&
 Lou 2008; Lou \& Shi 2013) as expected.
Clearly, $s\neq 0$ is the key difference brought
 by the Paczynski-Wiita gravity.
It greatly enriches the possible classes of
 self-similar dynamic solutions with event horizons
 and central expanding voids, as revealed by
 numerical solutions in Section 4.
% }

We shall presently discuss direct applications of these
 asymptotic analytical solutions at large enough $x$
 for numerical integrations of coupled nonlinear ODEs
 (\ref{ode1}) and (\ref{ode2}) using the standard
 Runge-Kutta shooting scheme.

\subsection{The absence of asymptotic analytic solutions
 for thermal expansion at large $x$}\label{thermal}

Wang \& Lou (2008) derived asymptotic self-similar
 solution for magnetohydrodynamic (MHD) thermal
 expansion at large $x$ (also Wang \& Lou 2007).
In this subsection, we naturally examine this
 possibility by removing the magnetic field and
 by invoking the Pasczinski-Wiita gravity in
 the form of expression (\ref{force}).
For such a thermal expansion, we expect $v(x)$
 to be proportional to $x$ as $x\rightarrow\infty$,
 namely
\begin{equation}
v(x)=bx+o(x)\ ,
\end{equation}
where $b$ is a constant proportional coefficient.
 Then nonlinear ODE (\ref{b1}) can be approximately written as
\begin{equation}
\frac{\mbox{d}\alpha}{\mbox{d}x}
+\frac{(2-3b)}{(1-b)}\frac{\alpha}{x}=0\ ,
\end{equation}
%checked to be correct!! June 9, 2012
giving the leading order term of $\alpha(x)$ as
\begin{equation}
\alpha=\alpha_e x^{-(2-3b)/(1-b)}+\cdots\ ,
\end{equation}
where $\alpha_e$ is a constant proportional coefficient. In order
 that the factor in the denominator $[x-s(x-v)x^2\alpha]$ of our
 nonlinear ODEs to be positive definite at large $x$, the leading
 order term of $\alpha$
 when $x\rightarrow\infty$ can at most be of power $-2$ for $x$.
This is also the condition for our approximate gravity force
 (\ref{force}) to be valid.
With this consideration, we require
\begin{equation}
%\frac{
(2-3b)/(1-b)\ge2\ .
\end{equation}
Here we take $b$ parameter to be non-zero and thus the above
 inequality demands $b\leq 0$, which would correspond to
 a homogeneous asymptotic collapse instead of an expansion.
However, the leading order of ODE (\ref{ode2}) then gives
\begin{equation}
(1-b)b=0\ ,
\end{equation}
indicating no such solutions can exist.
 The absence of thermal expansion or collapse solutions
 here is also due to the invalidity of the approximate
 gravity force under such conditions, since the outside
 gas mass density is not low enough.

%{\bf Stop here!}

%\subsection{Absence of asymptotic solutions for thermal expansion at large $x$}\label{thermal}
%For thermal expansion solutions, we expect $v(x)$ to be proportional to $x$ when $x\rightarrow\infty$, which is written down as
%\begin{equation}
%v(x)=bx+o(x)\ ,
%\end{equation}
%where $b$ is the constant coefficient. Then equation (\ref{b1}) can be approximated to be
%\begin{equation}
%\frac{2-3b}{1-b}\frac{\alpha}{x}+\frac{\mbox{d}\alpha}{\mbox{d}x}=0\ ,
%\end{equation}
%which gives the leading order term of $\alpha$
%\begin{equation}
%\alpha=\alpha_e x^{-\frac{2-3b}{1-b}}+\cdot\cdot\cdot\ ,
%\end{equation}
%
%
%where $\alpha_e$ is a constant. In order that the factor in the denominator $[x-s(x-v)x^2\alpha]$ in our ODEs to be positive definite at large $x$, the leading order term of $\alpha$ when $x\rightarrow\infty$ can at most be of power $-2$ of $x$. This is also the condition for our approximate gravitational force to be valid. With this limitation, we have
%\begin{equation}
%\frac{2-3b}{1-b}\ge2\ .
%\end{equation}
%Here we assume $b$ is non-zero, thus the above limitation demands $b<0$, which corresponds to a homogenous collapse instead of an expansion. However, the leading order of ODE (\ref{ode2}) then gives
%\begin{equation}
%(1-b)b=0\ ,
%\end{equation}
%which tells us such solutions can never exist, which agrees with the later numerical calculations. The absence of thermal expansion solutions is also due to the invalidity of our approximate gravitational force under such conditions, since the outside gas density is not low enough.

\subsection{Asymptotic analytic solutions
 of a diverging radial flow velocity $v(x)$
 at a finite $x$}

We here consider that $v(x)$ diverges
 at a finite value $x=x_1$; this point expands
 with time in a self-similar manner.
Then in the vicinity of $x_1$, $v(x)$
 becomes the dominant term, and nonlinear ODE
 (\ref{b1}) bears the local asymptotic form of
\begin{equation}
\frac{\mbox{d}(x^2\alpha)}{x^2\alpha}+\frac{\mbox{d}v}{v}=0\ ,
\end{equation}
immediately leading to
\begin{equation}
-x^2\alpha v\rightarrow m_0\ ,
\end{equation}
where $m_0$ is a positive constant of integration
 with the
 physical meaning of the enclosed mass within
 $x=x_1$ or the mass accretion rate there.
We presume that the divergent $v(x)$ behavior results from the
 approaching zero behavior of the denominator $[x-s(x-v)x^2\alpha]$
 in our nonlinear ODEs, which has the asymptotic form according
 to the above discussion (as will be shown later $\alpha$
 approaching zero there)
% {\bf Not precise!!}
\begin{equation}
x-s(x-v)x^2\alpha\rightarrow x-sm_0\ .
\end{equation}
We therefore identify the divergent point as $x_1=sm_0$.
 Physically, we have $s>0$.
Together with the condition of $\gamma\ge1$
 (which is almost always true), we verify that the
 ratio $\beta(x)/\alpha(x)$ remains finite at $x=sm_0$.
Consequently, the asymptotic form of nonlinear ODE
 (\ref{ode2}) becomes
\begin{equation}
-v\frac{\mbox{d}v}{\mbox{d}x}=\frac{m_0}{(x-sm_0)^2}\ .
\end{equation}
The corresponding asymptotic solution $v(x)$
 diverges near $x_1=sm_0$ in the leading-order form of
\begin{equation}\label{v-diverge}
v=-\bigg(\frac{2m_0}{x-sm_0}+v_d^2\bigg)^{1/2}\ ,
\end{equation}
where $v_d>0$ is a positive constant of integration.
 Here, $x_1=sm_0$ would correspond to the expanding
 boundary (event horizon) of a
 Schwarzschild black hole harbored at the center.
This implies in turn that the reduced mass density
 $\alpha(x)$ approaches zero fast enough in the
 vicinity of a Schwarzschild black hole boundary,
 namely
\begin{equation}\label{alpha_vanish}
\alpha=\bigg[s^2m_0\bigg(\frac{2m_0}{x-sm_0}
 +v_d^2\bigg)^{1/2}\bigg]^{-1}\ .
\end{equation}
Accordingly and as a result of sustained dynamic mass
 collapse, the
 dimensional radius $r_g$ of the Schwarzschild black hole
 with a mass $M$ grows linearly with time $t$ as
\begin{equation}\label{rg}
r_g=\frac{2GM}{c^2}=2ct\Big(\frac{s}{2}\Big)^{3/2}m_0\ .
\end{equation}
%{\bf Checked!}
In other words, the increasing rate of the
 Schwarzschild black hole radius is also
 directly related to $m_0$ parameter.

% {\bf Stop here!}

%{\bf
%We note that
In general relativity, the physical velocity is
 less than the speed of light $c$.
To incorporate this effect into the Paczynski-Wiita gravity,
 one should actually relate the true physical velocity
 $u_{tru}$ with the calculated velocity $u$ by
 $u=u_{tru}/(1-u_{tru}^2/c^2)$
%, where $c$ is the speed of light
 (Abramowicz et. al. 1996; Abramowicz 2009).
In this perspective, we interpret the divergence of $u$ near the
 black hole event horizon as the true velocity $u_{tru}$
 approaches $c$.
% }

\subsection{Asymptotic analytic solutions for the
  radial flow velocity approaching zero at small $x$}

%{\bf Lian Biao will add analysis on the asymptotic
% solution behaviors of enclosed mass related to
% the Schwarzschild radius. August 24, 2013. }
First, we study the case when $v(x)$ becomes negligible
 as compared with $x$ in the limit of $x\rightarrow 0$.
Among others, we presume $v(x)$ in the
 power-law form of
\begin{equation}\label{power}
v\rightarrow \Re \Big(v_sx^P\Big)\ ,
\end{equation}
%{\bf Real part operation added now!}
where $v_s$ is a complex constant proportional
 coefficient in general and $\Re(\cdots)$
 operation takes the real part of the
 argument;
while $P$ is a complex index with
 its real part $\Re (P)> 1$.
Conceptually, the imaginary part of $P$ would
 lead to an oscillatory behavior of $v(x)$.
Nonlinear ODE (\ref{b1}) then gives
\begin{equation}
\frac{1}{x^2\alpha}\frac{\mbox{d}(x^2\alpha)}
 {\mbox{d}x}=\Re\bigg(v_sPx^{P-2}\bigg)\ ,
\end{equation}
which readily leads to an asymptotic solution of
\begin{equation}
\alpha\approx\frac{\alpha_s}{x^2}
 \Big[1+\Re\Big(\frac{v_sP}{P-1}x^{P-1}\Big)\Big]\ ,
\end{equation}
%\begin{equation}
%\alpha
% =\frac{\alpha_s}{x^2}\exp\bigg(\frac{v_sP}
% {P-1}x^{P-1}\bigg)
% \approx\frac{\alpha_s}{x^2}
% \bigg[1+Re\bigg(\frac{v_sP}
% {P-1}x^{P-1}\bigg\bigg]\ ,
%\end{equation}
%%above 3 equations checked to be correct
where $\alpha_s$ is a positive constant proportional
 coefficient resulting from the analytic integration.
%{\bf
In this case, the mass $m(x)$ enclosed inside
 radius $x$ carries the asymptotic behavior of
 $m\rightarrow \alpha_s x$.
To ensure that the radius $x$ remains always larger
 than the dimensionless Schwarzschild radius $sm$
 where our theoretical model applies, we would
 require $s\alpha_s<1$.
% }
Substituting this approximate solution into nonlinear
 ODE (\ref{ode2}) and keeping up to the second-order
 terms, we derive the following relation
%\[\frac{\alpha_s}{(1-s\alpha_s)^2}\Big(1+\frac{1+s\alpha_s}
% {1-s\alpha_s}\frac{v_s}{P-1}x^{P-1}\Big)\]
%\begin{equation}\label{approx}
%\quad=2\alpha_s^{3\gamma-3}\Big\{ 1+\Big[(\gamma-1)
% \frac{2P+1}{P-1}-\gamma\frac{P}{2}\Big]v_sx^{P-1}\Big\}\ .
%\end{equation}
\[\frac{\alpha_s}{(1-s\alpha_s)^2}\Big\{1
 +\Re\Big[\Big(\frac{1+s\alpha_s}{1-s\alpha_s}
 \frac{1}{P-1}- (3-2\gamma)\Big)v_s x^{P-1}\Big]\Big\}\]
\begin{equation}\label{approx}
\quad=2\alpha_s^{3\gamma-3}\Big\{1+\Re\Big[\Big(
 \frac{3\gamma-3}{P-1}-\gamma-\frac{\gamma P}{2}
 \Big)v_sx^{P-1}\Big]\Big\}\ .
\end{equation}
Terms of the same powers of $x$ on both sides of relation
 (\ref{approx}) should be equal; we then have separately
 the following algebraic equations for determining $\alpha_s$
\begin{equation}
2\alpha_s^{3\gamma-4}(1-s\alpha_s)^2=1\ ,
\end{equation}
and for determining the complex power index $P$
\begin{equation}\label{P}
\gamma (P-1)^2+(7\gamma-6)(P-1)+\frac{2(1+s\alpha_s)}
{(1-s\alpha_s)}+6-6\gamma=0.
\end{equation}
%\begin{equation}\label{P}
%\gamma (P-1)^2-(3\gamma-4)(P-1)+4
%  +\frac{4}{1-s\alpha_s}-6\gamma=0\ .
%\end{equation}
%
%The above 2 equations checked to be correct.
%{\bf Stop here!}
Practically, we may regard $(P-1)$ as
 an unknown variable to be determined.
 This is then a quadratic equation for variable $(P-1)$.
We consider $\gamma\ge 1$, so $7\gamma-6>0$.
According to our presumption, we want to find a root
 with $\Re(P-1)>0$.
Such a root of $P$ is then impossible unless the constant
 term $2(1+s\alpha_s)/(1-s\alpha_s)+6-6\gamma<0$, namely
\begin{equation}
\gamma>1+\frac{(1+s\alpha_s)}{3(1-s\alpha_s)}\ .
\end{equation}
In the Newtonian gravity limit of $s\alpha_s\rightarrow 0$,
 this inequality is just $\gamma>4/3$.
%One notes that
This condition also indicates
 $P$ must be real, and so no oscillatory
 behavior would occur.

Another possible self-similar solution is that $v(x)$ is
 proportional to $x$ near the center, namely $v\rightarrow v_s x$
 with the proportional coefficient $v_s$ satisfying $0<v_s<1$.
To the leading order, nonlinear ODE (\ref{b1}) then gives
\begin{equation}
\alpha=\alpha_sx^{-2+v_s/(1-v_s)}\ .
\end{equation}
%{\bf
This indicates the reduced enclosed mass
 $m=(1-v_s)\alpha_sx^{v_s/(1-v_s)+1}$, dropping
  faster than $x$ as $x\rightarrow0$, so the
  dimensionless Schwarzschild radius $sm$
  remains smaller than $x$.
%Equation (\ref{ode2}) takes different limiting forms for different $\gamma$.
% }
Nonlinear ODE (\ref{ode2}) would be different
 for different regimes of $\gamma\neq 4/3$ values.
In the regime of $\gamma>4/3$, it gives
\begin{equation}
\frac{v_s}{(1-v_s)}=2\ ,\qquad\qquad v_s=2/3\ ,
\end{equation}
and
\begin{equation}
\alpha_s=v_s=2/3\ .
\end{equation}
%{\bf
On the other hand for $\gamma<4/3$, the second
 term on the right-hand side takes charge, and
% }
 yields
\begin{equation}
-\gamma v_s=2\gamma v_s-2\ ,\qquad v_s=2/(3\gamma)\ ,
\end{equation}
%{\bf
where $\alpha_s$ can take on any positive value.
%}
For these asymptotic behaviors discussed here, there
 will be no Schwarzschild black holes at the center.
%{\bf
For $\gamma>4/3$, the reduced mass density $\alpha(x)$
 approaches a constant $\alpha_s$ at $x=0$, while for
 $\gamma<4/3$, $\alpha(x)$ diverges as $x\rightarrow0$.
%  }
%For the regime of $\gamma<4/3$, it yields
%\begin{equation}
%-\gamma v_s=2\gamma v_s-2\ ,\qquad\qquad v_s=2/(3\gamma)\ ,
%\end{equation}
%with no constraints on the value of $\alpha_s$.
% For these asymptotic solution behaviors around $x=0$
% considered here, there is clearly no signature of a
% central Schwarzschild black hole in formation.
%{\bf Stop here.}

%Another possibility is that $v$ is proportional to $x$ near the center, $v\rightarrow v_sx$ with $0<v_s<1$. equation (\ref{b1}) then gives
%\begin{equation}
%\alpha=\alpha_sx^{v_s/(1-v_s)-2}\ ,
%\end{equation}
%while equation (\ref{ode2}) is different for different $\gamma$. For $\gamma>4/3$, it gives
%\begin{equation}
%\frac{v_s}{1-v_s}=2\ ,\qquad v_s=2/3\ ,
%\end{equation}
%and
%\begin{equation}
%\alpha_s=v_s=2/3\ .
%\end{equation}
%For $\gamma<4/3$, it yields
%\begin{equation}
%-\gamma v_s=2\gamma v_s-2\ ,\quad v_s=2/3\gamma\ ,
%\end{equation}
%while no constraint is on $\alpha_s$. For these asymptotic behaviors discussed here, there is no black hole at the center.
%}

\section{Construction of Numerical Self-similar Dynamic Solutions}

By properly specifying various analytic asymptotic solutions,
 we could construct numerical solutions from two coupled
 nonlinear ODEs (\ref{ode1}) and (\ref{ode2}) by
 straightforward numerical integrations.
We implement the fourth-order Runge-Kutta shooting scheme
 to integrate nonlinear ODEs (\ref{ode1}) and (\ref{ode2})
 numerically with proper initial conditions.
Then we plot the negative radial flow speed $-v(x)$ versus
 the independent self-similar variable $x$.
The critical curves of these two coupled nonlinear ODEs are
 found not to extend to infinity, instead they form closed
 loops when approaching large values of $x$, and sometimes
 they may even disappear, as shown below.
This is an interesting novel feature of introducing
 the Paczynski-Wiita gravity in our
 self-similar hydrodynamic analysis.
The behaviors of the solutions in the vicinity of critical
 curves are detailed in Appendix A.

%\section{Numerical Solutions}

%Beginning with various asymptotic solutions we could construct numerical solutions of ODEs (\ref{ode1}) and (\ref{ode2}). We employ the fourth-order Runge-Kutta method to integrate ODEs (\ref{ode1}) and (\ref{ode2}) numerically with proper initial conditions. Then we plot $-v(x)$ with respect to the self-similar radius variable $x$. The critical lines of the ODEs are found not to extend to infinity, instead they form closed loops when approaching large $x$, and sometimes they even disappear. This will be discussed below in section \ref{sect-cri}. The behavior of the solutions in the vicinity of critical lines are analyzed in Appendix A.

\subsection{Critical curves of the nonlinear ODEs and
 expansion-wave collapse solutions (EWCSs)}\label{sect-cri}

When the coefficients of $\mbox{d}v/\mbox{d}x$ and
 $\mbox{d}\alpha/\mbox{d}x$ on the left-hand side (LHS)
 of nonlinear ODEs (\ref{ode1}) and (\ref{ode2}) become
 zero, the right-hand side (RHS) must vanish simultaneously
 in order to ensure finite values of first derivatives
 $\mbox{d}v/\mbox{d}x$ and $\mbox{d}\alpha/\mbox{d}x$.
% {\bf Consistent for both RHSs?}
These two conditions give the equation for the sonic
 critical curve of the two coupled nonlinear ODEs, viz.
\begin{equation}
\frac{(x-v)}{x}-\frac{\gamma-1}{\gamma}
 =\frac{\gamma^{\frac{1}{3-3\gamma}}
 (x-v)^{\frac{\gamma+1}{3\gamma-3}-1}x^{2/3}/2}
 {\big[x-s\gamma^{\frac{1}{3-3\gamma}}
 (x-v)^{\frac{\gamma+1}{3\gamma-3}}x^{2/3}\big]^2}\
\end{equation}
 (see Appendix B for detailed derivations).
For a necessary consistency, this result has been
 carefully checked against the pertinent results
 of Wang \& Lou (2008).
Sonic critical lines carry the physical meaning that
 they separate between supersonic and subsonic flows.
It is obvious that for $\gamma>1$, $x-v=0$
 is a branch of the critical line.
This line is special in the sense that it cannot be
 crossed, since we must require a positive enclosed
 mass $m=x^2\alpha(x-v)$ everywhere.
Therefore, this straight line is referred to as
 the zero-mass line (ZML) for which the enclosed
 mass remains zero.
% {\bf Waiting for more analysis from Biao
% ...email April 12, 2012}
For self-similar dynamic solutions touching the ZML,
 expanding voids emerge around the center and evolve
 with time $t$ in a self-similar manner.
Physically, a central energy supply (e.g. photon gas,
 high-energy neutrinos and/or electron-positron
 $e^{\mp}$ pair plasma) would be required to initiate
 and sustain such void formation and evolution (e.g.
 Lou \& Wang 2011).

%\subsection{Critical lines of the ODEs and expansion-wave collapse solutions}\label{sect-cri}
%When the coefficients in front of $\mbox{d}v/\mbox{d}x$ or $\mbox{d}\alpha/\mbox{d}x$ on the left hand side of ODEs (\ref{ode1}) and (\ref{ode2}) vanish, the right hand side must also vanish in order to prevent $\mbox{d}v/\mbox{d}x$ or $\mbox{d}\alpha/\mbox{d}x$ from a divergence. This requirement gives the equation for the critical line of the ODEs:
%\begin{equation}
%\frac{(x-v)^2}{x}-\frac{\gamma-1}{\gamma}(x-v)=\frac{\gamma^{\frac{1}{3-3\gamma}}(x-v)^{\frac{\gamma+1}{3\gamma-3}}x^{2/3}} {2[x-s\gamma^{\frac{1}{3-3\gamma}}(x-v)^{\frac{\gamma+1}{3\gamma-3}}x^{2/3}]^2}\ .
%\end{equation}
%Critical lines have the physical meaning that they separate between supersonic and subsonic flows. It is obvious that for $\gamma>1$, $x-v=0$ is one of the critical lines, which we shall call the first branch. This line is special in the sense that it cannot be crossed, since the enclosed mass $m=x^2\alpha(x-v)$ must be non-negative everywhere. As we will show later, solutions may end on this critical line.

\begin{figure}
\includegraphics[width=88mm]{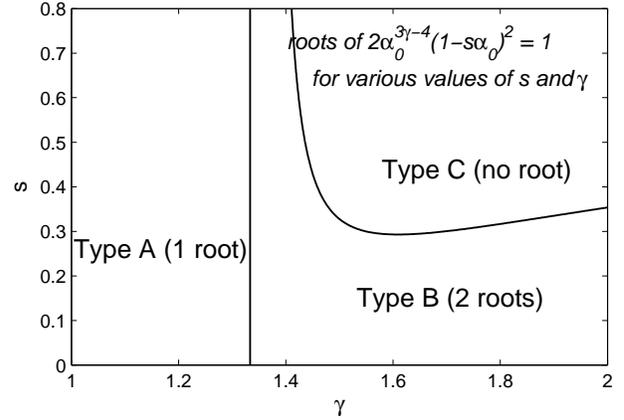}
%\caption{Different types of parameters $s$ and $\gamma$, classified according to the number of roots of equation $2\alpha_0^{3\gamma-4}(1-s\alpha_0)^2=1$ (see Appendix B). Each root $\alpha_0$ corresponds to an expansion-wave collapse solution with $\alpha=\alpha_0/x^2$ and $v=0$ in large $x$ limit. For type C region, no critical lines exist.}
\caption{Different regimes for parameter pair
 $s$ and $\gamma$, classified according to the number
 of $\alpha_0$ roots of equation
 $2\alpha_0^{3\gamma-4}(1-s\alpha_0)^2=1$
 (see Appendix B).
The solid vertical line marks $\gamma=4/3$.
 To the left of this $\gamma=4/3$ vertical line (i.e.,
 type A regime), there exists one root for $\alpha_0$.
To the right of this $\gamma=4/3$ vertical line, the
 solid curve separates the no root regime (i.e., type C)
 from the two root regime (i.e., type B) as indicated.
Right along the $\gamma=4/3$ vertical line, we would have
% two roots of $\alpha_0=0$ and $\alpha_0=2/s$, although
 two roots of $\alpha_0=(1\pm 1/\sqrt{2})/s$
 (the plus sign root is unphysical), although
 the case of $\gamma=4/3$ is not considered in this paper
 (see Goldreich \& Weber 1980 and Lou \& Cao 2008 for
 $\gamma=4/3$ with the Newtonian gravity).
Each root $\alpha_0$ may correspond to an expansion-wave
 collapse solution (EWCS) with $\alpha=\alpha_0/x^2$ and
 $v=0$ in the regime of very large $x$.
In the type C regime, no sonic critical
 lines exist.}\label{fig:pcri}
\end{figure}

%{\bf Stop here!}

The behavior of the second branch of the critical lines is
 found closely related to the number of roots of $\alpha_0$
 from the following equation
\begin{equation}
2\alpha_0^{3\gamma-4}(1-s\alpha_0)^2=1\ . \label{alpha0'}
\end{equation}
%{\bf What is the origin of this critical line equation?}
 (see equations \ref{alf} and \ref{18}).
The number of roots of equation (\ref{alpha0'}) for various
 pair values of $s$ and $\gamma$ is shown in Figure \ref{fig:pcri}
 (see Appendix B for a detailed analysis).
In the regime of $\gamma<4/3$, there is always one root only
 for $\alpha_0$, and we refer to this regime as Type A.
In the regime of $\gamma>4/3$, there exists a curved boundary
 separating the ``two roots" regime (referred to as Type B)
 and the ``no root" regime (referred to as Type C).
Note that for $s<(1-\sqrt{2}/2)=0.2929$ and $\gamma>4/3$,
 there are always two roots for $\alpha_0$.
%The deeper mathematical reason of this connection is beyond our discussion, so we only present the result here. The number of roots of equation (\ref{alpha0'}) for various values of $s$ and $\gamma$ is shown in Figure \ref{fig:pcri} (see Appendix B for details). In the region $\gamma<4/3$, there is always only one root, and we denote this region as Type A. In the region $\gamma>4/3$, there is a boundary between "two roots" region(denoted as Type B) and "no root" region(denoted as Type C). Note that for $s<(1-\sqrt{2}/2)=0.2929$ and $\gamma>4/3$, there are always two roots.

\begin{figure}
\includegraphics[width=88mm]{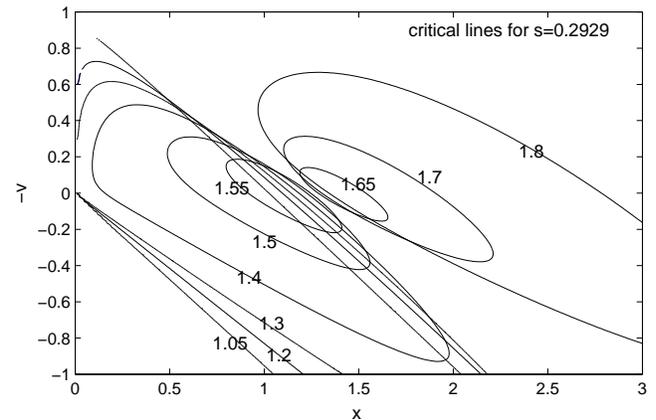}
\caption{Sonic critical curves for different $\gamma$
 values marked along the curves all with the same
 value for $s=0.2929$.
For $\gamma<4/3$, the critical curves touch the
 origin $x=0,\ v=0$; three examples are shown for
 $\gamma=1.05,\ 1.2,\ 1.3$, respectively.
For $\gamma>4/3$ ranging from $1.4$ to $1.8$, the sonic
 critical lines form loops which shrink when the
 parameters approach the type C regime shown in
 Figure \ref{fig:pcri} (around $\gamma\sim 1.6$).}
%\caption{Critical lines for different $\gamma$ with $s=0.2929$. For $\gamma<4/3$, the critical line touches the origin $x=0,v=0$. For $\gamma>4/3$, the critical lines form a loop, and shrink when the parameters get close to type C region shown in figure \ref{fig:pcri}.}
 \label{fig:cri}
\end{figure}

%{\bf Stop here!}

The
%second branch of the
 sonic critical curve in general
 appears like a closed loop.
In Figure \ref{fig:cri}, we show several such
 sonic critical curves for $s=0.2929$ with different
 $\gamma$ values ranging from $1.05$ to $1.8$; the
 straight ZML $x-v=0$ is not shown in this figure.
In the Type A regime of $\gamma<4/3$, the sonic
 critical loop curve touches
 the origin for the cases of $\gamma=1.05,\ 1.2,\ 1.3$,
 respectively, and it seems that the loop ``breaks" into two branches
 in the vicinity of the origin (e.g., see the sonic critical
 curve of $\gamma=1.2$ in Figure \ref{fig:cri} and also see
 the dash-dotted curve in Figure \ref{fig:gamma-1.2}).
It can be shown that the sonic critical curve touches the
 origin in the manner $v=x/\gamma$ (this would differ
 from the ZML for $\gamma>1$ with different slopes).
In the Type B regime, the critical curve forms a closed loop,
 and is away from the origin, as shown in Figure \ref{fig:cri}
 (see also the dash-dotted curves in Figures \ref{fig:gamma-1.7}
 and \ref{fig:gamma-1.4a}).
In the Type C regime, it is found that the sonic critical curve
%(second branch)
 disappears completely.
In Figure \ref{fig:cri}, we observe how the closed sonic critical
 loop moves step by step when the parameter pair $(\gamma,\ s)$
 gradually approaches the Type C regime.

%The second branch of the critical line is in general a closed loop like curve. In Figure \ref{fig:cri} we show some critical lines for $s=0.2929$ and different $\gamma$ from $1.2$ to $1.8$(the first branch $x-v=0$ omitted). In the Type A region($\gamma<4/3$), the loop curve touches the origin and it seems that the upper half of the loop is squeezed onto the vertical axis in the vicinity of the origin(see the critical line $\gamma=1.2$ in Figure \ref{fig:cri} for instance, see also the dashed-dotted line in Figure \ref{fig:gamma-1.2}). It can be shown that the lower half of the loop touches the origin in the manner $v=x/\gamma$. In the Type B region, the loop leaves apart from the origin, as shown in Figure \ref{fig:cri}(also the dashed-dotted line in Figures \ref{fig:gamma-1.7} and \ref{fig:gamma-1.4a}). In the Type C region, it is found that the second branch critical line disappears. In Figure \ref{fig:cri} we can see how the critical closed loop shrinks step by step when the point $(\gamma, s)$ get closer to the Type C region.

This behavior of the sonic critical curve becomes better
 understood as we explore numerically \emph{expansion-wave
 collapse solutions} (EWCS), as first studied by \citet{Shu1977}
 for an isothermal gas under the Newtonian self-gravity (see
 also further extensions by Lou \& Shen 2004 and references
 therein).
An EWCS contains an expansion wave front outside of which the gas
 remains static (in an equilibrium between the gas pressure and
 the self-gravity) and inside of which the gas gradually collapses
 towards a central free-fall to form a mass-accreting core,
 which would be a black hole with an event horizon in the
 present context.
%{\bf
Basically, it describes a spontaneous collapse of
 a general static singular polytropic sphere (SPS)
 solution starting with a power-law mass density
 profile and zero velocity, i.e.
% \[v=0\ ,\]
\begin{equation}\label{singular}
v(x)=0\ ,\qquad\qquad\qquad \alpha(x)=\frac{\alpha_0}{x^2}\ ,
\end{equation}
where $\alpha_0$ is the root of equation (\ref{alpha0'}).
A singular static gas sphere (be it isothermal,
 conventional polytropic or GP) results from the balance
 of pressure force and
 self-gravity and is generally considered as
 unstable in both the non-relativistic limit (e.g.,
 Shu 1977; Lou \& Shen 2004) and the relativistically
 hot or degenerate gas with $\gamma=4/3$ (e.g., Cai
 \& Shu 2003).
Once an infinitely small expansion-wave front emerges
 from the central core region, the affected sphere will
 continuously collapse in a self-similar dynamic manner.
% }
The expansion wave front always lies on the sonic critical
 curve (see figure 2 of Shu 1977 for an isothermal gas;
 see also Tsai \& Hsu 1995).
It is then apparent that in the Type C regime where no
 $\alpha_0$ root of equation (\ref{alpha0'}) exists,
 there should be no EWCSs accordingly.
%expansion-wave collapse solution.
In other words, there is no sonic critical curve
 available for the expansion wave front to sit at.
Now we turn to Type B regime, where the sonic critical
 curve forms a closed loop away from the origin in a $-v$
 versus $x$ figure presentation.
There are two roots of $\alpha_0$ available for constructing
 the outside SPS solutions, and a numerical
 integration does give two possible EWCSs with two different
 event horizons.
The two corresponding Schwarzschild radii increase with
 constant yet different speeds for the same $s$ but
 different $m_0$ values.
%  expansion-wave collapse solutions.
This is a novel feature of forming black holes in
 a static mass reservoir stretched to larger radii.
In both Figures \ref{fig:gamma-1.7} and \ref{fig:gamma-1.4a}, the
 two EWCSs in each case are displayed by heavy solid curves.
As is easily seen, different choices of $\alpha_0$ give
 distinct locations of the expansion wave front.
The loop shaped sonic critical curve cuts the horizontal $x$ axis
 twice, and is exactly able to hold two expansion wave fronts,
 corresponding to the two roots of $\alpha_0$ respectively.
The outer (or faster) EWCS seems to be more ``dense" than the
 inner one,
 in the sense that it corresponds to a higher $\alpha_0$ value.
In Type A regime, however, there is only one root of
 $\alpha_0$, and we thus expect only one EWCS,
 which is numerically verified indeed.
In Figure \ref{fig:gamma-1.2} for $\gamma=1.2$,
 such a solution is shown in heavy solid curve.
It is also easy to understand this
 result from the perspective of the sonic critical curve.
In this situation the sonic critical loop breaks,
 and is connected to the origin.
So it cuts the positive $x$ axis only once,
 and can only hold one expansion wave front.
Based on the discussion above, we see that there exists
 a one-to-one correspondence between the behavior of
 the sonic critical curve and the emergence of EWCSs
 with corresponding event horizons.

We note in particular that EWCSs shown above are quite
 different when the Paczynski-Wiita gravity
%general relativity
 is considered for the dynamic collapse of a GP gas sphere.
%{\bf
We expect these EWCSs to be generally valid in the
 exact general relativity formalism, since the
 Paczynski-Wiita gravity closely approximates
 Einstein's general relativity in describing the strong-field deviation of the gravitational
 potential from the $1/r$ law.
Particularly, the discussion of EWCSs based on the
 number of roots of equation (\ref{singular}) is
 sufficiently general, and should depend little on
 the details of gravitational potential.
% }
Cai \& Shu (2005) considered gravitational collapses
 in the general relativistic formalism for singular
 isothermal spheres (SIS) with $\gamma=1$, and found
 only one EWCS.
Our conclusions here are consistent with their results.
 More importantly, we emphasize that for $\gamma>4/3$ there can be
 two (for a wide range of proper combinations of $s<1$ and $\gamma$
 parameters including the range of a sufficiently small $s$)
 or no (for sufficiently large $s<1$ and $\gamma$ values) EWCSs.
%{\bf
In the case of two EWCSs, we identify the one with smaller
 $\alpha_0$ as the counterpart of the usual EWCS in the Newtonian
 limit, while the second one with larger $\alpha_0$ as a novel
 solution describing the collapse of a highly compact general
 SPS with strong non-Newtonian self-gravity.
%Physically, the second EWCS at low $s$ represents the collapse of a highly dense singular general polytropic sphere balancing its interior pressure by strong non-Newtonian self-gravitation.
For spherical gas clouds with larger $s<1$ values, the absence
 of EWCSs is due to the absence of general SPSs under the
 predominantly strong non-Newtonian self-gravity.
We note here that the existence of two or no general SPSs for
 $\gamma>4/3$ is similar to the result of Oppenheimer \&
 Volkoff (1939) on the neutron star problem, except that
 their physical system has a finite boundary.
In their general relativistic formalism which has effectively
 $4/3<\gamma\le5/3$, the number of equilibrium solutions runs
 from one (non-relativistic) to two (non-relativistic and
 relativistic), and then to zero as the compact
 stellar mass increases.
% }
It would be very interesting to extend the self-similar
 general relativistic analysis of Cai \& Shu (2005) for
 a GP gas sphere and examine the number of possible
 EWCSs with event horizons.
%For gas clouds with large $s$ values, the absence of
% such solutions may be due to the inability for the
% gas to support itself with too high a mass density.
For massive spherical clouds with an ineffective heat
 transfer, the polytropic index $\gamma$ can be
 physically larger than $4/3$.
We expect
%{\bf
 such analysis will give two or no EWCSs
 when the formalism of general relativity is completely
 taken into account.
% }
%our conclusion to remain valid when the formalism of
% general relativity is fully taken into account.

%It should be remarked that expansion-wave collapse solutions shown above are quite different when general relativity is considered for collapse of polytropic spheres. In the work of Cai \& Shu (2005) gravitational collapse under general relativity is considered for singular isothermal spheres (SIS) which equivalently have $\gamma=1$, and they have found one expansion-wave collapse solution. Our conclusions here agree with theirs, and we find for $\gamma>4/3$ there may be two (for small $s$) or no (for $s$ large enough) expansion-wave solutions. For gas clouds with a large $s$, the absence of such solutions may be due to the inability for the gas to support itself with too high densities. For cloud spheres with a low heat transfer, the polytropic $\gamma$ is possibly larger than $4/3$. We expect this conclusion will stay correct when general relativity is seriously taken into account.

Approaching the Newtonian gravity limit of $s\rightarrow 0$,
 the behaviors of the sonic critical curve differ
 for $\gamma<4/3$ and $\gamma>4/3$, respectively.
For the case of $\gamma<4/3$, the sonic critical
 curve does not change much.
For the case of $\gamma>4/3$, the sonic critical
 curve (closed loop) enlarges itself, and the outer
 side of the loop extends to infinity gradually.
Thus in the extreme Newtonian gravity limit of $s=0$, only
 one EWCS remains inside a finite radius $r$ for any values
 of $\gamma$.

%{\bf Stop here.}

%When approaching the Newtonian limit $s\rightarrow 0$, the behavior of the critical line are different for $\gamma<4/3$ and $\gamma>4/3$. For $\gamma<4/3$ the critical line does not change much. For $\gamma>4/3$, the second branch critical line (loop) enlarges itself, and the upright side of the loop extends to infinity. Thus in the Newtonian limit $s=0$, only one expansion-wave collapse solution remains at finite radius for any value of $\gamma$. Thus, the existence of the additional expansion-wave collapse solution for $\gamma>4/3$ is a pure general relativity effect.

\subsection{Other continuous self-similar solutions}

By specifying proper values of $v_0$ and $\alpha_0$ and
 integrating inwards from a certain sufficiently large
 $x$, various continuous dynamic self-similar solutions
 can be constructed.

%\subsection{Other continuous solutions}
%By proposing proper values of $v_0$ and $\alpha_0$ and integrating inward from some large $x$, various continuous solutions can be constructed. We select some representative examples and present them below.

%In Figure \ref{fig:s-4}, the parameters $(\gamma, s)$ are in the Type C region, and the loop critical line vanishes. The function $-v(x)$ of a group of solutions are plotted with respect to $x$ in the figure. The dotted vertical lines represent the boundary of the central black hole, where the function $v$ diverges according to equation (\ref{v-diverge}). $\alpha_0$ is fixed to be $0.180$, while different solutions correspond to different value of $v_0$. Generally speaking, a higher velocity of outflow $v_0$ corresponds to a smaller sized black hole. When the velocity $v_0$ is large enough, there is not a black hole but a void at the center, where both pressure and density of the gas vanishes. The slope of $v(x)$ of the void solution when $x$ approaches the critical line $x-v=0$ is well predicted in Appendix \ref{A}.

\begin{figure}
\includegraphics[width=88mm]{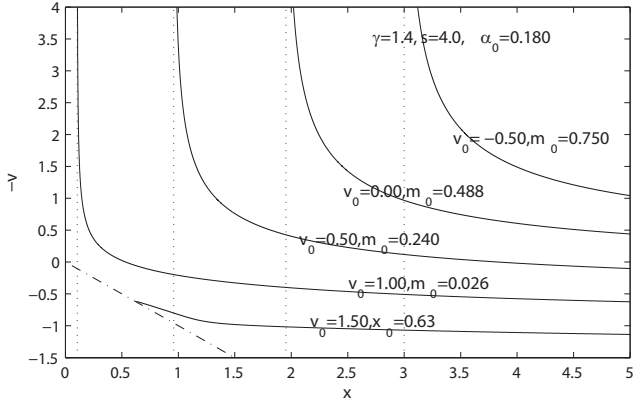}
%\caption{Numerical solutions with $\gamma=1.4$ and $s=4.0$ (type C region). Dotted vertical lines indicate the boundary of the black hole in each solution. Neither critical lines nor expansion-wave solutions exist. The lowest solution has a void core instead of a black hole in the center.}
\caption{Numerical self-similar dynamic solutions with
 $\gamma=1.4$ and $s=4.0$ in the type C regime.
Four
%{\bf Need to add the 4th dotted line in the figure;
% email sent on Aug 21, 2013; done Aug 27, 2013}
 dotted vertical lines indicate the boundaries
 of central black holes in each solution.
The corresponding masses and radii of central black
 holes increase with time in a self-similar manner.
Neither sonic critical curves nor EWCSs exist.
 The lowest solution touching the straight dash-dotted
 ZML $v=x$ at $x_0=0.63$ has a central spherical void
 in dynamic self-similar expansion instead of a growing
 central black hole.
%This void expands with time in a self-similar manner.
%{\bf Possible solutions with $v_0<0$?}
Expanding void solutions can be constructed in sequence by
 further increasing the value of velocity parameter $v_0$.
In general, numerical solutions with $v_0>0$ (outflow),
 $v_0=0$ (static), and $v_0<0$ (inflow) are all possible.
 }\label{fig:s-4}
\end{figure}

%{\bf Stop here!}

%Both Figure \ref{fig:gamma-1.7} and Figure \ref{fig:gamma-1.4a} show several solutions for $(\gamma, s)$ in the Type B region. In Figure \ref{fig:gamma-1.7} it is assumed that $\gamma=1.7$ and $s=0.2$, while in Figure \ref{fig:gamma-1.4a} $\gamma=1.4$ and $s=0.2929$. The two expansion-wave collapse solutions are shown in heavy solid lines. Generally speaking, there are two main branches of solutions. One branch of solutions do not cross the second branch critical line, including the outer expansion-wave solution, as shown in Figure \ref{fig:gamma-1.4a} in light solid lines. The other branch of solutions cross the second branch critical line twice, and the inner expansion-wave solution belongs to this kind. The light solid lines in Figure \ref{fig:gamma-1.7} correspond to this branch of solutions. They usually have a larger value of $\alpha_0$ and a smaller center black hole than the former branch of solutions when their $v_0$s are comparable to each other.

\begin{figure}
\includegraphics[width=88mm]{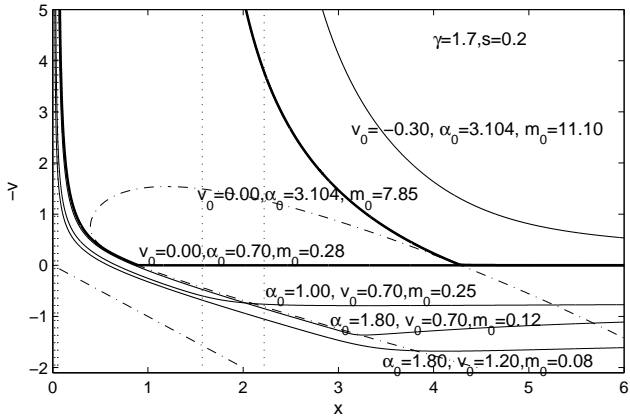}
%\caption{Numerical solutions with $\gamma=1.7$ and $s=0.2$ (type B region). The dashed-dotted line is the critical line. The two black solid lines are two expansion-wave collapse solutions. Some solutions crossing the critical line twice are shown, which usually have a smaller $\alpha_0$.}
\caption{Numerical self-similar dynamic solutions with
 $\gamma=1.7$ and $s=0.2$ in the type B regime.
The dash-dotted curve is the sonic critical curve.
The straight dash-dotted line at the lower left
 corner is the ZML.
 The two heavy solid curves are the two possible EWCSs.
Several self-similar solutions crossing the sonic critical
 line twice are also shown; they usually involve smaller
 $\alpha_0$ values.
The dotted straight vertical lines mark the
 expanding radii of the growing central black holes.
Key pertinent parameters are marked along solution curves.
It is possible to construct dynamic self-similar solutions
 with $v_0>0$, $v_0=0$ and $v_0<0$ far away.
%{\bf Black holes in this Figure 4?
% Corresponding $m_0$ values?}
% {\bf Possible solutions with $v_0<0$?}
  }\label{fig:gamma-1.7}
\end{figure}

\begin{figure}
\includegraphics[width=88mm]{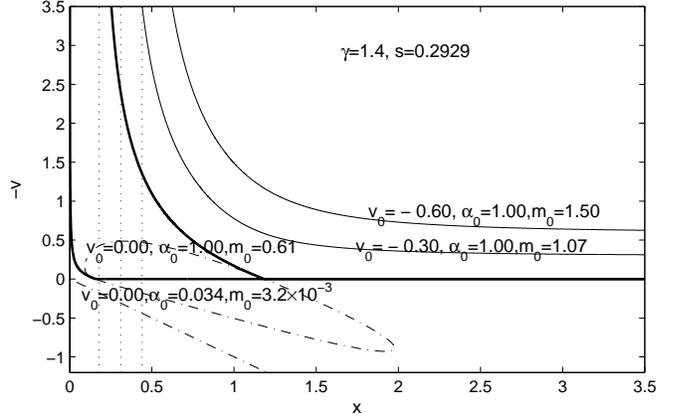}
%\caption{Numerical solutions with $\gamma=1.4$ and $s=0.2929$ (type B region). The dashed-dotted line is the critical line. The two black solid lines are the two expansion-wave collapse solutions. Several solutions not crossing the critical line are shown, which in general have a larger $\alpha_0$.}
\caption{Numerical self-similar dynamic solutions with
 $\gamma=1.4$ and $s=0.2929$ in the type B regime.
The dash-dotted loop is the sonic critical curve.
The straight dash-dotted line at the lower left
 corner is the ZML.
The two heavy solid curves are the two possible EWCSs.
%expansion-wave collapse solutions.
Two self-similar solutions (both with $v_0<0$) not
 crossing the sonic critical curve are also shown;
 in general, they have larger $\alpha_0$ values.
Dotted vertical lines on the left side indicate the
 expanding boundaries of central black holes.
% {\bf Need four $m_0$ values.}
Pertinent parameters are marked along the solution curves.
 }\label{fig:gamma-1.4a}
\end{figure}

\begin{figure}
\includegraphics[width=88mm]{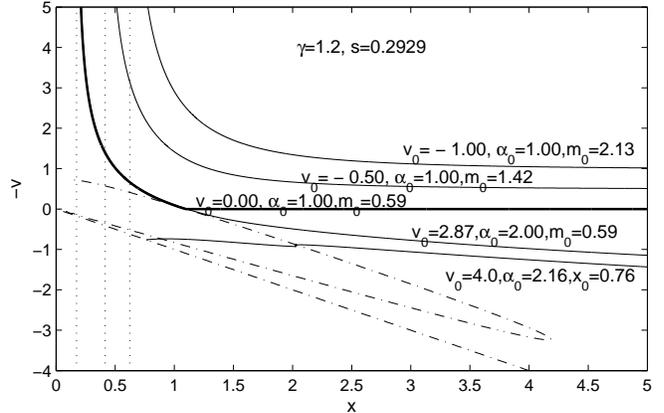}
%\caption{Numerical solutions with $\gamma=1.2$ and $s=0.2929$ (type A region). The dashed-dotted line is the critical line. The black solid line is the only one expansion-wave collapse solution. The lowest solution covalence with the expansion-wave collapse solution inside the expansion-wave front.}
\caption{Numerical self-similar solutions with
 $\gamma=1.2$ and $s=0.2929$ in the type A regime.
The partial dash-dotted elongated loop broken
 near $x=0$ is the sonic critical curve.
The straight dash-dotted line at the
 lower left corner is the ZML.
 The heavy solid curve is the only EWCS.
The lowest solution coincides
%covalence
 with the EWCS inside
 the expansion-wave front.
The dotted straight vertical lines indicate the
 expanding radii of the growing central black holes.
Key pertinent parameters are marked
 along the solution curves.
It is also possible to construct a void solution
 that passes across the elongated sonic critical
 loop (broken at small $x$) twice smoothly.
%{\bf Central black holes and their parameters?}
% {\bf Possible void solutions?}
% {\bf Can they pass across the elongated critical
% loop twice smoothly?}
 }\label{fig:gamma-1.2}
\end{figure}

In Figure \ref{fig:s-4}, the two parameters $\{\gamma,\ s\}$
 are chosen from the Type C regime, and thus the sonic
 critical loop does not appear.
The function for negative reduced speed $-v(x)$ versus $x$
 for a group of continuous self-similar solutions are
 numerically obtained and displayed in Figure \ref{fig:s-4}.
The dotted straight vertical lines represent the expanding
 boundaries (event horizons) of the central Schwarzschild
 black holes, where the magnitude of the function $-v(x)$
 diverges to infinity according to asymptotic analytic
 solution (\ref{v-diverge}).
Here, mass parameter $\alpha_0$ is fixed at a given value of
 $0.180$, while different numerical solutions correspond to
 different values of velocity parameter $v_0$ far away.
The corresponding values of $m_0$ are also indicated as well.
 Generally speaking, a higher outflow velocity $v_0$ far away
 corresponds to a smaller size for the central black hole.
When the velocity parameter $v_0$ becomes large enough,
 there is no central black hole but instead an expanding
 void emerges at the center, where both pressure and mass
 density of the gas vanish.
In astrophysics, a central energy source (or a ``fire ball")
 is needed in order to carve out and retain such a dynamically
 expanding void into a massive envelope or shell.
In various pertinent contexts, we have recently explored several
 astrophysical applications of such dynamical voids in
 self-similar expansions (e.g., Lou \& Wang 2012; Lou \& Zhai
 2009, 2010; Lou \& Hu 2010).
With conceptual consistency, dynamical self-similar void in
 expansions can also emerge with the Paczynski-Wiita gravity.
The slope of $v(x)$ of the void solution when $x$ approaches the
 ZML $x-v=0$ is well determined analytically in Appendix \ref{A}.

In both Figures \ref{fig:gamma-1.7} and
 \ref{fig:gamma-1.4a}, we show several constructed
 solutions for chosen $\{\gamma,\ s\}$ pairs in the Type B regime.
In Figure \ref{fig:gamma-1.7}, pertinent parameters are
 $\gamma=1.7>5/3$ and $s=0.2$, while in Figure \ref{fig:gamma-1.4a},
 we specify the parameter pair $\gamma=1.4>4/3$ and $s=0.2929$.
The two possible EWCSs with central event horizons
 are shown by the heavy solid curves.
Generally speaking, there are two main branches of solutions. One
 branch of solutions does not cross the oval sonic critical curve,
 including the outer expansion-wave solution, as shown in
 Figure \ref{fig:gamma-1.4a} by the light solid curves.
The other branch of solutions crosses the oval sonic critical
 curve twice, and the inner expansion-wave solution
 belongs to this kind.
The light solid curves in Figure \ref{fig:gamma-1.7}
 correspond to this branch of solutions.
They usually have a larger value of mass parameter $\alpha_0$
 and a smaller or less massive central black hole than the
 former branch of solutions when their velocity parameter
 $v_0$ values are comparable to each other.

Figure \ref{fig:gamma-1.2} is plotted to demonstrate
 self-similar solutions of two coupled nonlinear ODEs
 (\ref{ode1}) and (\ref{ode2}) in the Type A regime.
We take $\gamma=1.2$ and $s=0.2929$. As is easily seen,
 the elongated sonic critical curve breaks in
 the vicinity of the origin.
The only EWCS with central event horizon
 is shown by the heavy solid curve.
 Several other solutions with different $v_0$ values
 are displayed in light solid curves above the EWCS.
The lowest solution in this figure is connected smoothly
 onto the EWCS at the expansion-wave front on the
 sonic critical curve.
These two solutions correspond to two different derivatives
 $\mbox{d}v/\mbox{d}x$ across the sonic critical curve.
The analysis of the two possible first derivatives
 $\mbox{d}v/\mbox{d}x$ across the sonic critical
 curve can be found in Appendix \ref{A}.

%Figure \ref{fig:gamma-1.2} is plotted to demonstrate solutions of the ODEs in the Type A region. We assume $\gamma=1.2$ and $s=0.2929$. As is easily seen, the second branch critical line is connected to the origin. The only one expansion-wave collapse solution is plotted in heavy solid line. Several other solutions with different values of $v_0$ are shown in light solid lines above the expansion-wave collapse solution. The lowest solution in the figure is connected onto the expansion-wave collapse solution at the expansion-wave front on the critical line. These two solutions correspond to two different derivatives $\mbox{d}v/\mbox{d}x$ on the critical line. The analysis of the two possible derivatives $\mbox{d}v/\mbox{d}x$ near the critical line may be found in Appendix \ref{A}.

\begin{figure}
\includegraphics[width=88mm]{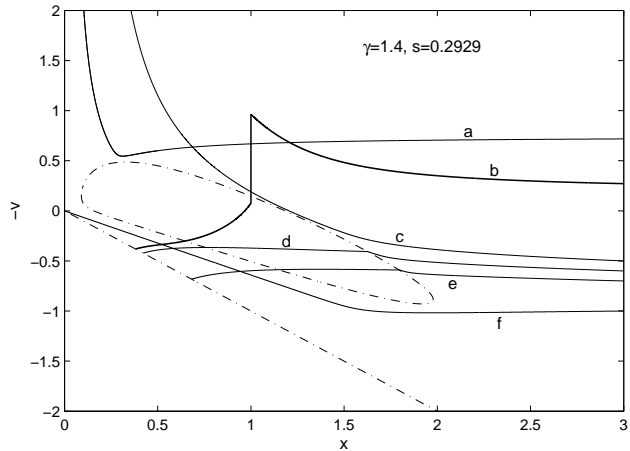}
%\caption{Several solutions with $\gamma=1.4$ and $s=0.2929$. The dashed-dotted line is the critical line. Solutions $d$ and $e$ have a discontinuity in $\mbox{d}v/\mbox{d}x$ when crossing the critical line. Solution $b$ is connected by a shock wave near the critical line. Solution $f$ smoothly reaches the center without singularities.}
\caption{Shown here are several examples
 of self-similar dynamic solutions with
 $\gamma=1.4$ and $s=0.2929$.
The closed dash-dotted curve is the sonic critical
 loop and the straight dash-dotted line represents
 the zero mass line (ZML).
We obtain these numerical solutions by specifying
 $\alpha_0$ and $v_0$ at a sufficiently large $x$
 and integrating towards small $x$.
For solutions $a$ and $c$, there are event horizons
 for black holes at the center of collapse.
Solutions $d$ and $e$ have discontinuities in
 the first derivative $\mbox{d}v/\mbox{d}x$
 when crossing the sonic critical loop,
%  {\bf Twice?}
 and have expanding voids inside the gas cloud.
Solutions $c$, $d$ and $e$ have similar behaviors
 at large $x$, but differ when encountering the
 sonic critical curve.
Solution $b$ involves a shock wave
 across the sonic critical loop
% {\bf smoothly sonic critical loop?}
 and a central expanding void.
%Solution $b$ is connected by a shock wave which
% crosses the critical line, inside which a void region presents.
Solution $f$ smoothly reaches the center $x=0$ very
 closely without singularity, and is obtained by
 carefully tuning the initial $\alpha$ and $v$ at large $x$.
If one tunes the initial velocity $v$ larger (smaller) while
 keeping the initial $\alpha$ invariant, the integrated
 solution will develop an expanding void (black hole) at
 the center.
Thus, $f$ solution appears to separate solutions with
 central voids and solutions with central black
 holes.
%  {\bf Meaning?}
Solutions $b$, $d$, and $e$ have central voids
 expanding outwards in self-similar manners.
% {\bf Descriptions for solutions $a$ and $c$?
% Solutions $a$ and $c$ involve central black holes.} --
%{\bf Several solutions with $\gamma=1.4$ and $s=0.2929$. The dashed-dotted lines are the critical lines. We obtain the solutions by assuming $\alpha$ and $v$ at some large $x$ and integrating towards small $x$. For solutions $a$ and $c$, there are black holes at the center of collapse. Solutions $d$ and $e$ have a discontinuity in $\mbox{d}v/\mbox{d}x$ when crossing the critical line, and have voids inside the gas cloud. Solutions $c$, $d$ and $e$ have similar behaviors at large $x$, but differs when encountering the critical line. Solution $b$ is connected by a shock wave which crosses the critical line, inside which a void region presents. Solution $f$ smoothly reaches the center without singularities, and is obtained by carefully tuning the initial $\alpha$ and $v$ at large $x$. If one tunes the initial velocity $v$ larger (smaller) while keeps the initial $\alpha$ invariant, the integrated solution will develop a void (black hole) at the center.}
 }\label{fig:gamma-1.4b}
\end{figure}

%Discontinuity in $\mbox{d}v/\mbox{d}x$ may arise when solutions are crossing the critical lines. This is due to the indefiniteness of $\mbox{d}v/\mbox{d}x$ on the critical line (Appendix \ref{A}). Different derivatives $\mbox{d}v/\mbox{d}x$ may lead to distinct behaviors inside the core. Explicit examples are given in Figures \ref{fig:gamma-1.4b} and \ref{fig:7shock}. Compared to solution $c$ in figure \ref{fig:gamma-1.4b} which has a black hole in the center, solution $d$ and $e$ cross the critical line with a jump in $\mbox{d}v/\mbox{d}x$, leading to a void core. Solution $c$ and $d$ in figure \ref{fig:7shock} have the same small $x$ curve, but separate after hitting the critical line. Solution $f$ in figure \ref{fig:gamma-1.4b} and solution $b$ in figure \ref{fig:7shock} are connected to zero at $x=0$, with neither black hole nor void core in the center.

Discontinuities in the first derivative $\mbox{d}v/\mbox{d}x$
 may arise when solutions cross the sonic critical curves
 (e.g. Whitworth \& Summers 1985).
% {\bf See reference of Whitworth \& Summers on weak
% discontinuities in the isothermal cases! Two pertinent
% references added and need to include relevant discussion.
% Two emails sent to Biao in this regard.}
Different values of the velocity first derivative
 $\mbox{d}v/\mbox{d}x$ may lead to distinct behaviors
 around the central collapsing core.
A few explicit examples are shown in Figures
 \ref{fig:gamma-1.4b} and \ref{fig:7shock}.
% ---
% {\bf
In Figure \ref{fig:gamma-1.4b}, solutions $c$,
  $d$ and $e$ are constructed by integrating inwards
  from the same yet sufficiently large $x$, with the
  same initial reduced density $\alpha_0$ but slightly
  different initial reduced velocity $v_0$.
Compared to solution $c$ in Figure \ref{fig:gamma-1.4b}
  which has a black hole (event horizon)
  at the center, solutions $d$
  and $e$ cross the sonic critical curve with a jump
  in $\mbox{d}v/\mbox{d}x$, leading to corresponding
  voids around the center in self-similar expansions.
They nevertheless have similar dynamic behaviors at
  sufficiently large $x$.
In Figure \ref{fig:7shock}, solutions $c$, $d$ and $e$
 are obtained by integrating outwards from small $x$.
They have the same behavior before hitting the sonic
 critical curve.
Solution $c$ does not actually go across the
 sonic critical curve, and has a discontinuity in
 $\mbox{d}v/\mbox{d}x$ when touching the sonic
 critical curve; this is implemented by adopting
 the two eigenvalues of $\mbox{d}v/\mbox{d}x$ on
 the sonic critical curve.
Solution $d$ crosses the sonic critical curves twice,
 and has a discontinuity in $\mbox{d}v/\mbox{d}x$ when
 hitting the sonic critical curve for the second time.
Solution $e$ features a shock wave solution and a
 central black hole with an expanding event horizon
 which will be discussed further in the next subsection.

Solution $f$ in Figure \ref{fig:gamma-1.4b} and solution
 $b$ in Figure \ref{fig:7shock} are examples of solutions
 reaching the origin at $x=0$, with neither black holes
 nor voids around the center.
In terms of solution structures, they appear to
 represent a kind of gradual transition from
 central voids to central Schwarzschild black holes.
These solutions are obtained by carefully tuning
 the initial values of $\alpha_0$ and $v_0$ for
 inward integration at a sufficiently large $x$.
Starting from solution $f$ in Figure \ref{fig:gamma-1.4b}
 as an example, if one tunes the initial velocity $v_0$ to
 be larger (smaller) while keeping the initial $\alpha_0$
 unchanged, the integrated numerical solution will give
 rise to a void (or black hole) around the center.
Solution $b$ has a discontinuity in $\mbox{d}v/\mbox{d}x$
 when crossing the sonic critical curve, and if one tunes
 the initial velocity $v_0$ to be larger (or smaller)
 while keeping the initial $\alpha_0$ unchanged, the
 integrated numerical solution leads to a void.
% (or encounter unphysical numerical results with non-zero
% imaginary value) {\bf Meaning?} at the center. }
% ---
%Compared with solution $c$ of Figure \ref{fig:gamma-1.4b} which
% has a black hole anchored at the center, solutions $d$ and $e$
% cross the sonic critical line with a jump in the first
% derivative $\mbox{d}v/\mbox{d}x$, leading to
% corresponding central voids in self-similar expansions.
%Solution $b$ in Figure \ref{fig:7shock} and solution $f$
% in Figure \ref{fig:gamma-1.4b} reach the origin $x=0$,
% with neither a black hole nor a void at the center.
%In terms of solution structures, they represent a kind
% of gradual transition from central voids to central
% Schwarzschild black holes.

The reduced mass density functions $\alpha(x)$ of numerical
 solutions in Figure \ref{fig:7shock} are plotted in
 Figure \ref{fig:density} in correspondence.
% ---
% {\bf
We use the same labels $a$, $b$, $c$, $d$, $e$ as
 those of their corresponding reduced velocities $v(x)$
 shown in Figure \ref{fig:7shock}.
For solutions $a$, $c$, $d$, $e$ with black holes at
 the center, the mass density drops to zero
 rapidly when approaching the Schwarzschild radius
 (i.e., event horizon).
This is an important effect due to a rapid
 acceleration the gas experienced near the
 event horizon of a black hole.
For solution $b$ which continues to the origin $x=0$,
 the mass density diverges at the center, as
 concluded in Section 3.4 for $\gamma<4/3$.
At large radii, the reduced mass density
 functions all drops with a $1/x^2$ scaling.
% }
% ---
%We use the same labels $a$, $c$, $d$, $e$ as those
% of their reduced velocities $v(x)$ plotted in
% Figure \ref{fig:7shock}.
%As is easily seen, the density drops to zero rapidly
% when approaching the Schwarzschild radius.
%This is the effect of the drastic acceleration the gas
% experienced near the black hole boundary.
%At large distances, the density function drops in
% a $x^{-2}$ scaling behavior.
When a solution hits the sonic critical curve, there is a
 discontinuity in $\mbox{d}\alpha/\mbox{d}x$ in association
 with the discontinuity in $\mbox{d}v/\mbox{d}x$.

%\subsection{Shock waves}
\subsection{Dynamic shock waves in self-similar expansion}

It is possible to construct self-similar dynamic
 solutions with expanding shocks.
Across a shock wave front, discontinuities in gas mass density,
 radial flow velocity, gas pressure and specific entropy occur.
By setting the constant proportional coefficient ${\cal C}=1$
 with no loss of generality in equation (\ref{beta1}) for the
 reduced form of GP EoS with $\gamma\neq 4/3$,
 we need to assign different sound parameters $k_+$ and $k_-$
 for the upstream and downstream sides of a shock front.
Accordingly, we use subscript indices $+$ and
 $-$ to distinguish variables at upstream and
 downstream sides of a shock, respectively.
%{\bf
In principle for shock conditions in the framework
 of reference co-moving with the shock, we require
 conservations of mass, radial momentum and energy
 across a shock.
% }
More specifically, the jump conditions for shock
 wave solutions can be obtained from coupled nonlinear
 ODEs (\ref{b1}), (\ref{ode1}) and (\ref{ode2})
 immediately by integration from the left-hand side
 (downstream) of the discontinuous point to the
 right-hand side (upstream).
%The connecting conditions for shock wave solutions consist of continuity of the mass flow density, change of the momentum flow density by pressure difference and continuity of the energy flow density, in the shock wave front's static reference frame.
%
In terms of reduced functions $\alpha(x)$
 and $v(x)$, we can cast these jump
 conditions across the shock as
\begin{equation}
k_-^{1/2}x_-=k_+^{1/2}x_+\ ,
\end{equation}
\begin{equation}
k_-^{3/2}(x_--v_-)x_-^2\alpha_-
 =k_+^{3/2}(x_+-v_+)x_+^2\alpha_+\ ,
\end{equation}
\begin{equation}
k_-[\alpha_-(x_--v_-)^2+\beta_-]
 =k_+[\alpha_+(x_+-v_+)^2+\beta_+]\ ,
\end{equation}
\[k_-^{3/2}\Big[\frac{\alpha_-}{2}(x_--v_-)^3
 +\frac{\gamma\beta_-(x_--v_-)}{(\gamma-1)}\Big]\]
\begin{equation}
\qquad=k_+^{3/2}\Big[\frac{\alpha_+}{2}(x_+-v_+)^3
 +\frac{\gamma\beta_+(x_+-v_+)}{(\gamma-1)}\Big]\ ,
\end{equation}
where the upstream ($+$) and downstream
 ($-$) reduced dimensionless pressures
 $\beta_\pm=\alpha_\pm^{3\gamma-2}
 (x_\pm-v_\pm)^{2(\gamma-1)}x_\pm^{4(\gamma-1)}$
 are defined at the shock front.
% Checked against Wang & Lou (2008) August 22, 2013
For an isothermal gas of $\gamma=1$, we
 replace the term $\beta/[\alpha(\gamma-1)]$ in
% {\bf Need a confirmation! Aug 22, 2013;
% Confirmed by Skype conversation with Lian Biao
% ShanXi TaiYuan August 23, 2013 afternoon}
 the equations by its limit $\ln [\alpha^3(x-v)^2x^4]$.
The law of thermodynamics for the increase of
 specific entropy from the upstream side to
 the downstream side further requires the
 flow velocity relative to the shock must
 be supersonic on the upstream side and
 subsonic on the downstream side.

%{\bf Stop here!}

%--------Solving the above equations:
%\[A=\frac{(x_+-v_+)\alpha_+}{x_+}\ ,\]
%\[B=\frac{1}{\alpha_+}+A^{2\gamma-4}x_+^{6\gamma-8}\alpha_+^\gamma\ ,\]
%\[C=\frac{1}{\alpha_+^2}+\frac{2\gamma}{\gamma-1}A^{2\gamma-4}x_+^{6\gamma-8}\alpha_+^{\gamma-1}\ ,\]
%\[\frac{\gamma+1}{\gamma-1}\Big(\frac{1}{\alpha_-}\Big)^2-\frac{2\gamma B}{\gamma-1}\Big(\frac{1}{\alpha_-}\Big)+C=0\ ,\]
%\[\alpha_+\alpha_-=\frac{\gamma+1}{\gamma-1}\frac{1}{C}\ .\]
%--------

%The connecting conditions for shock wave solutions can be obtained from equations (\ref{b1}), (\ref{ode1}) and (\ref{ode2}) immediately by integration from the left side of the discontinuous point to the right side. Assume the solution is discontinuous at $x=x_c$, then the connecting conditions are as follows:
%\begin{equation}\label{conne1}
%\alpha^+(x_c-v^+)=\alpha^-(x_c-v^-)=K\ ,
%\end{equation}
%\begin{equation}\label{conne2}
%\frac{\gamma}{\gamma-1}K^{3\gamma-3}x_c^{4\gamma-4}(x_c-v)^{1-\gamma}+\frac{1}{2}(x_c-v)^2\Big|^{x_c^+}_{x_c^-}\ ,
%\end{equation}

%where $K$ is an arbitrary constant, $v^\pm$ and $\alpha^\pm$ corresponds to the left and right limits of $v$ and $\alpha$ at $x=x_c$. When $\gamma=1$, one should use $\ln(x_c-v)$ to replace $(x_c-v)^{1-\gamma}/(1-\gamma)$ in equation (\ref{conne2}). With this condition for connection, shock waves may be obtained intermediately.

\begin{figure}
\includegraphics[width=88mm]{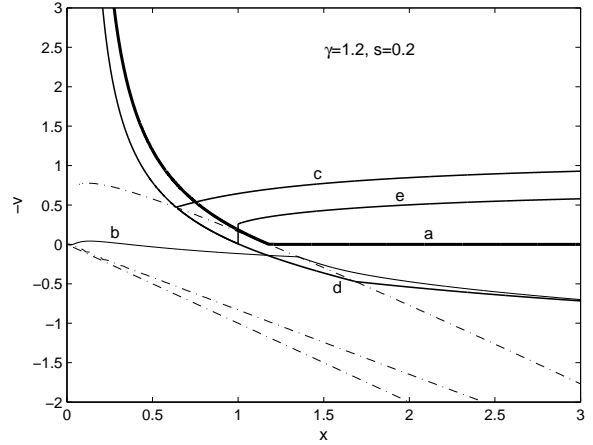}
\caption{Numerical solutions with $\gamma=1.2$ and $s=0.2$.
 The dash-dotted curve broken in two segments is the
  sonic critical curve, while
% {\bf
 the straight dash-dotted line is the ZML.
% Confirmed with Lian Biao Aug 23, 2013 afternoon Shan Xi.
Solution $a$ in boldface line is the EWCS with an event
 horizon of central black hole.
Both solutions $b$ and $c$ have discontinuities in
 $\mbox{d}v/\mbox{d}x$ when crossing the sonic
 critical curve.
Solution $b$ smoothly reaches $x=0,\ v=0$, and has a discontinuity
 in $\mbox{d}v/\mbox{d}x$ when crossing the critical curve.
It is obtained by integrating inwards with carefully chosen
 initial values of $\alpha_0$ and $v_0$ at a sufficiently large $x$.
%Solution $b$ is smoothly connected to $x=0,v=0$ without
% singularities.
Solutions $c$, $d$ and $e$ are obtained by integrating
 outwards from small $x$.
Two eigenvalues for the first derivative $\mbox{d}v/\mbox{d}x$
 are allowed at the sonic critical curve.
Solution $c$ does not cross the critical curve smoothly and
 has a discontinuous $\mbox{d}v/\mbox{d}x$ there.
% when hitting the sonic critical curve.
Solution $d$ crosses the sonic critical curve twice.
Solution $e$ is obtained by creating a shock across
 the sonic critical curve.
This is an example of a black hole inside an
 expanding shock.
%Solution $e$ is involved with an event horizon to a central
% black hole via a shock wave.
% {\bf How about d? will add more descriptions and explanations;
% transition between voids and event horizons; Skype with
% Lian Biao in ShanXi August 23, 2013 afternoon.}
% ---
%{\bf Numerical solutions with $\gamma=1.2$ and $s=0.2$. The dashed-dotted lines represent the critical lines. Solution $a$ is the expansion-wave collapse solution, which we denote by a solid line. Solution $b$ smoothly reaches $x=0,v=0$, and has a discontinuity in $\mbox{d}v/\mbox{d}x$ when crossing the critical line. It is obtained by integrating inwards with carefully chosen inital $\alpha$ and $v$ at a large $x$. Solution $c$, $d$ and $e$ are obtained by integrating outwards from small $x$. Two values of $\mbox{d}v/\mbox{d}x$ are allowed at the critical line. Solution $c$ does not cross the critical line, and has a discontinuous $\mbox{d}v/\mbox{d}x$ when hitting the critical line. Solution $d$ cross the critical line twice. Solution $e$ is obtained by creating a shock across the critical line. This is an example which has a black hole inside the shock wave front.}
 }\label{fig:7shock}
\end{figure}

\begin{figure}
\includegraphics[width=88mm]{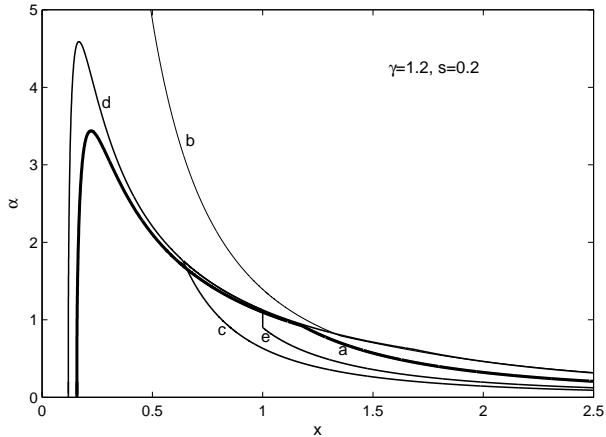}
\caption{Reduced mass density function $\alpha(x)$
 corresponding to numerical solutions for $v(x)$ in
 Figure \ref{fig:7shock} with the parameter pair
 $\gamma=1.2$ and $s=0.2$.
% {\bf Where is $b$ solution? The corresponding
% $\alpha(x)$ divergent towards to the origin; Lian Biao
% will add the b curve with descriptions and explanations;
% transition between voids and event horizons; Skype with
% Lian Biao in ShanXi August 23, 2013 afternoon. Done
% August 27, 2013.}
The labels $a$, $b$, $c$, $d$, $e$ denote the reduced
 mass density function of the corresponding velocity
 solutions
 denoted by the same labels in Figure \ref{fig:7shock}.
At the Schwarzschild radius, the reduced mass density
 $\alpha(x)$ drops
 to zero, as shown in equation (\ref{alpha_vanish}).
Discontinuities in $\mbox{d}\alpha/\mbox{d}x$ occur
 in accompany with the corresponding discontinuities
 in $\mbox{d}v/\mbox{d}x$ when the solutions cross
 the sonic critical curve.
For solution $b$ where there is no void or black hole at
 the center, the mass density diverges as $x\rightarrow 0$;
 it represents a kind of transition between voids and
 event horizons of central black holes.
For the shock wave solution $e$, the reduced mass
 density function $\alpha(x)$ is discontinuous at
 the shock.
%{\bf Density function $\alpha$ of numerical solutions in figure \ref{fig:7shock} with $\gamma=1.2$ and $s=0.2$. The letters $a$, $b$, $c$, $d$, $e$ denote the density function of the corresponding solutions denoted by the same letter in figure \ref{fig:7shock}. At the Schwarzschild radius, the density $\alpha$ drops to zero, as demonstrated in equation (\ref{alpha_vanish}). Discontinuities in $\mbox{d}\alpha/\mbox{d}x$ occur in accompany with the siscontinuities in $\mbox{d}v/\mbox{d}x$, when the solutions hit the critical line. For solution $b$ where there is no void or black hole at the center, the density diverges as $x\rightarrow0$. For the shock wave solution $e$, the density function $\alpha$ is discontinuous across the shock.}
 }\label{fig:density}
\end{figure}

Examples of self-similar shock solutions are given
 as solution $b$ in Figure \ref{fig:gamma-1.4b}
 and as solution $e$ in Figure \ref{fig:7shock}.
In order to avoid the ambiguity brought by using different
 sound parameters $k_\pm$, we multiplied the downstream
 $v_-$ and $x_-$ in the presentation of solution $b$ in
 Figure \ref{fig:gamma-1.4b} by a factor $(k_-/k_+)^{1/2}$
 (the upstream side of the shock naturally involves $k_+$),
 while we multiplied the upstream $v_+$ and $x_+$
 in the presentation of solution $e$ in Figure
 \ref{fig:7shock} by a factor $(k_+/k_-)^{1/2}$ (the
 downstream side of the shock naturally involves $k_-$).
% {\bf Need to be specific and precise! Carefully
%checked with Lian Biao at ShanXi, August 23, 2013}
The reduced mass density function $\alpha(x)$ of
 the shock solution $e$ in Figure \ref{fig:7shock}
 is shown in Figure \ref{fig:density} in
 correspondence.

We note that there is a void instead of a black
 hole at the center of solution $b$ in Figure
 \ref{fig:gamma-1.4b},
 shrouded within the shock wave front.
%In such a self-similar accretion solution, one might
% have expected a black hole at the center by analysing
% the large radius gas inflow, which turns out to be
% not the truth.
Physically, this may imply that the central shock
 wave inside a progenitor during a supernova
 explosion may delay the formation of a black hole
 and/or other compact remnants, supporting a
 transient void at the center of explosion
 (Lou \& Wang 2012).
Such a void could be filled with electron/positron
 $e^{\mp}$ pair plasma, high-energy neutrinos,
 and intense radiation field.
When the shock weakens and peters out, a black
 hole or a compact remnant may or may not form under
 the stellar self-gravity in a fall-back process,
 depending on how fast the massive envelope expands.

In principle by our formalism with the Paczynski-Wiita gravity,
 one could further explore many more abundant solutions with
 various weak discontinuities across the sonic critical curve,
 for instance, double-shock solutions, as has been done in some
 earlier works within the Newtonian gravity formalism (e.g.,
 Whitworth \& Summers 1985; Lazarus 1981).
These solutions may be used to describe specific physical
 processes associated with mass collapses to form
 black holes, which will be pursued in separate papers.

\section{Physical self-similar solutions
 and realistic black hole systems}

The possible values for the two parameters $s$ and $\alpha_0$
 to adopt may take the following physical considerations into
 account.
By referring to self-similar transformation (\ref{self-similar}),
 one readily derives the following exxpression
\begin{equation}\label{s}
s=\frac{2(p/\rho)}{c^2\alpha^{3\gamma-3}
 (x-v)^{2\gamma-2}x^{4\gamma-4}}\ ,
\end{equation}
where $p$ and $\rho$ are the gas pressure and
 mass density, while $c$ is the speed of
 light in vacuum.
As a simple way to estimate the above expression, we
 take the limit of $x\rightarrow\infty$ (corresponding
 to $t\rightarrow 0$ for a fixed $r$ or
 $r\rightarrow\infty$ at a fixed $t$), where
 $\alpha\rightarrow\alpha_0/x^2$ and
 $v\rightarrow $ constant.
If we regard $p/\rho$ approximately as the square
 of the sound speed denoted by $u_S^2$, we have a
 simple estimation
\begin{equation}
s\alpha_0^{3\gamma-3}=2\frac{u_S^2}{c^2}\ .
\end{equation}
Another equation that can be easily derived
 from self-similar transformation
 (\ref{self-similar}) is
\begin{equation}\label{eq46}
\frac{GM}{r}=k\frac{m(x)}{x}\ .
\end{equation}
We then regard $GM/r$ as the square of the Keplerian
 velocity $u_K^2$ and thus for large $x$,
 equation (\ref{eq46}) above yields
\begin{equation}
s\alpha_0=2\frac{u_K^2}{c^2}\ ,
\end{equation}
which is usually much smaller than $1$.
 At the event horizon or boundary of a
 black hole, we have $GM/r_g=c^2/2$ and
 $x=sm_0$ where $m_0$ is the reduced
 dimensionless black hole mass.
After a time $t$ of sustained self-similar
 mass accretion and collapse, the radius
 $r_g$ for the event
 horizon of a black hole increases linearly
 with time $t$ according to equation
 (\ref{rg}) as $r_g=s^{3/2}m_0ct/2^{1/2}$.
Naturally, another physical requirement is
 simply $s^{3/2}m_0<2^{1/2}$ such that the
 expansion speed of Schwarzschild radius
 $r_g$ should remain less than $c$.

%In a typical accretion process of a black hole inside
% a galaxy core, the gas temperature can be as high as
% $\sim 10^8$ K or even higher.
%Keplerian velocity $u_K$ and sound speed $u_S$ are of
% the same order (due to Virial theorem) around $\sim
% 10^2\ {\rm km\ s}^{-1}$ to $\sim 10^3\ {\rm km\ s}^{-1}$.
%Such dynamic accretion processes may be possibly represented
% by an EWCS with a parameter range of $s\sim 10^{-7}$ to
% $\sim 10^{-5}$ and $\alpha_0\sim 1$ (In the limit $s\alpha_0\rightarrow 0$, it is easily seen from equation (\ref{alpha0'}) that $\alpha_0\sim 1$.). The physical meaning of parameter $s$ is then the square of the ratio of sound speed $u_S$ (or Keplerian velocity $u_K$) and light speed $c$ in large radius limit. These expansion wave solutions usually has a reduced black hole mass $m_0\sim0.1$ to $1$. Thus a super massive black hole with $10^9$ solar mass, or equivalently with a radius of $10^{10}km$, can be quickly formed through accretion in $10^5$ to $10^7$ years. This growth time is rather short compared to the age of galaxies (about $10^{10}$ years), because spherical accretions with large effective accreting area are much more efficient than disk accretions. Such behaviors are similar to Bondi accretion (Bondi 1952; Bondi \& Hoyle 1944), which also basically describes accretions of spherical gas clouds.

%{\bf Where did you get the following info?}
In plausible mass collapse and accretion processes
 for a black hole inside a galactic bulge or
 central region in a cluster of galaxies,
% {\bf Clusters of galaxies?}
 the gas temperature can be as high as
 $\sim 10^7$ K to $\sim 10^8$ K or even higher.
The Keplerian velocity $u_K$ and the sound speed
 $u_S$ are more or less in the same order of
 magnitudes (due to the Virial theorem) around
 $\sim 10^2$ km s$^{-1}$ to $\sim 10^3$ km s$^{-1}$.
Such dynamic mass collapse and accretion processes may
 be possibly represented by an EWCS with a parameter
 range of $s\sim 10^{-7}$ to $\sim 10^{-5}$ and
 $\alpha_0\sim 1$; in the limit of
 $s\alpha_0\rightarrow 0$, it is easily seen from
 condition (\ref{alpha0'}) for an EWCS with an event
 horizon that $\alpha_0\sim 1$.
The physical meaning of parameter $s$ is the square of
 the ratio of the sound speed $u_S$ (or comparably, the
 Keplerian velocity $u_K$)
 to $c$ in the regime of very large radius $r$
 (i.e. sufficiently far away from an accreting central
 supermassive black hole -- SMBH).
By our sample calculations, these EWCSs usually have
 a reduced black hole mass $m_0\sim 0.1$ to $1$.
Therefore a SMBH with a mass of $\sim 10^9$ M$_{\odot}$,
 or equivalently with a Schwarzschild radius of
 $\sim 3\times 10^{9}$ km, may rapidly emerge through
 a quasi-spherical dynamic mass collapse process in a
 range of timescales from $\sim 10^5$ to $\sim 10^7$ years
 with an important proviso for the presence of sustained
 mass reservoirs surrounding the central singularity
 of spacetime.
%This growth time is rather short compared to the age of
% typical galaxies ($\sim 10^{10}$ yrs), because spherical
% dynamical accretions with large effective accreting
% area are much more efficient than disk accretions.
In other words for cosmological contexts, it is
 conceivable to find rare hypermassive black holes (HMBHs)
 in an even higher mass range of $\sim 10^{10}$ M$_{\odot}$
 to $\sim 10^{12}$ M$_{\odot}$ as long as hypermassive
 matter reservoirs (e.g. rich clusters of galaxies with
 $\sim 10^{15}M_{\odot}$) could be formed and maintained
 in the universe and they could sustain quasi-spherical
 dynamic mass collapses towards central SMBHs/HMBHs.
Our estimated SMBH growth time is fairly short as compared
 to the age of typical galaxies ($\sim 10^{10}$ yrs), mainly
 because quasi-spherical dynamic collapses with large
 effective accreting areas are much more efficient than
 disk accretions due to the well-known difficulty of
 removing disk angular momentum.
Such behaviors are similar to the familiar Bondi accretion
 (Bondi 1952; Bondi \& Hoyle 1944), which nevertheless
 describes a steady-state spherical mass accretion of
 a conventional polytropic gas onto a gravitating
 central mass point (i.e., the gas self-gravity
 is ignored in the well-known Bondi model).

Quasi-spherical dynamic mass collapses might indeed dominate
 in the early stages of black hole formation and growth, when a
 seed black hole (or a singularity) remains small and surrounded
 by a gas cloud, or a remnant supermassive star (SMS), that
 is sufficiently dense and in large amount (e.g. Hoyle \& Fowler
 1963; Fowler 1964; Begelman 2009).
Under such favorable conditions, a black hole forms and accretes
 from an almost spherically symmetric massive envelope or
 reservoir, and mass accretions are effective in all directions.
It might be plausible to suspect that SMBHs in most galaxies
 probably might have experienced an early fast growth phase
 when conditions are favorable in the very early epochs of
 the universe well before the formation of disk structures
 of galaxies, and powerful quasars with high red shifts $z>$
 6 and 7 or even more (e.g. Mortlock et. al. 2011) and AGNs
 may appear during such unusually fast mass collapses
 and accumulations.
Early universe and later epochs may allow cold dark matters
 (DMs) to form loosely bound systems or pre-galaxies under
 the self-gravity (e.g. Peebles 1982; Blumenthal et. al.
 1984), which are most probable to take a grossly
 spherical structure.
Such quasi-spherical DM structures may be present in
 the universe ubiquitously without being detected by
 a large fraction so far.
They become more easily detected when they dynamically
 interact with normal matters in gas phase leading to
 active radiations.
As an example, nearby DM dominated dwarf galaxy Segue 1
 may reveal an extremely important clue along this line
 of reasoning (e.g. Xiang-Gr\"uss, Lou, \& Duschl 2009).
Some extremely high mass density peaks may also exist in
 the early universe according to numerical simulations,
 due to density fluctuations and sound waves left after
 the extremely hot and dense initial cosmological stages
 (e.g. Tolman 1934; Ford \& Parker 1977; Lukash 1980;
 Mukhanova et al. 1992; Lyth \& Riotto 1999).
Such fluctuating spacetime may allow relatively local
 quasi-spherical mass collapse processes of early
 black holes, and SMBHs with masses exceeding
 $\sim 10^9$ M$_{\odot}$ or $\sim 10^{10}$ M$_{\odot}$
 (i.e. HMBHs) are possibly formed in several million
 years (e.g. Hu, Shen, Lou \& Zhang 2006).
Over the past several decades, growing evidences show the
 extremely high red shift bright quasars and many AGNs are most
 likely powered by accreting SMBHs (e.g. Begelman et al. 1984;
 Rees 1984; Kormendy \& Richstone 1995
% {\bf ; Fan X.H. et al.?; and more updated references}
 and extensive references therein), while the rapid formation
 of extremely massive compact objects within such a short
 timescale in the early universe is still not well understood.
Through the above physical estimates and analyses, we propose
 that when quasi-spherical dynamic collapses and accretions
 are available with enough supporting mass reservoirs, the
 growth rates of SMBHs may be able to reach a sensible level.

%{\bf Stop here！}

%{\bf
We briefly compare our theoretical model analysis
  with the simulations on general relativistic collapses
  of the supermassive stars (SMS) (Hoyle \& Fowler 1963)
  up to a mass limit of $\sim 10^6 M_{\odot}$ (e.g. Shibata
  \& Shapiro 2002; Linke et. al. 2001; Saijo \& Hawke 2009;
  Montero et. al. 2012), which are strikingly similar to
  ours here.
All simulations show a rapid increase of the inflow
 velocity when approaching the black hole event horizons,
 which agrees with our conclusion qualitatively.
In most simulations, the velocity as a function of radius indeed
 shows certain invariant patterns with increasing time, but not
 necessarily self-similar.
The absence of self-similar collapses in most simulations
 may be related to an initial global perturbation (a sudden
 reduction of global pressure), which might not be physical.
We suspect that for perturbations triggered at the central
 region, collapses in a self-similar form may be very
 likely to emerge.
In a realistic astrophysical system, the pressure
 instability usually starts from the center.
We expect future simulations based on such central
 local perturbations to test this scenario.
We note that the timescale of forming a seed SMBH
 in the simulations is only $10\sim 10^2$ yrs,
 much shorter than that estimated here.
The reason is that a SMS is a far denser and more
 compact object (supporting nuclear reactions)
 than a spherical cloud that we consider here;
 a SMS is much faster to collapse below the
 Schwarzschild radius.

Under more violent conditions of forming stellar mass
 black holes during supernova explosions, the central core
 temperature may reach $\sim 10^{11}$ K to $\sim 10^{12}$ K
 or even higher, and the surrounding gas envelope may become
 extremely relativistic.
The sound speed $u_S$ may approach the relativistic upper bound
 of $c/3^{1/2}$ (e.g. Weinberg 1972), and the Keplerian velocity
 $u_K$ may also be comparable to $c$.
With these estimates, the value of $s$ parameter may lie within
 the range of $0.1\sim 1$, while $\alpha_0\sim 1$ with
 $\alpha=\alpha_0/x^2$ scaling in the regime of large $x$.
If the emergence of central stellar-mass black holes under
 these conditions are estimated by our self-similar dynamic
 collapse solutions, the timescale of a complete
 formation would be within $\sim 10^{-3}$ s.
This short dynamic timescale is one to two orders of magnitude
 less than that of the supernova explosion process including
 stellar core collapse and the emergence of a powerful rebound shock.
This dynamic timescale appears reasonable and suggestive, though a
 realistic supernova explosion process would be much more involved
 to model general relativistic space-time structures
 (Oppenheimer \& Snyder 1939; Tolman 1939; Misner \& Sharp 1964,
 1965; Misner 1965), very high-energy nuclear physics (Bethe 1990
 and references therein) and various hydrodynamic instabilities
 (Blondin et al. 2003; Buras et al. 2006; Arnett \& Meakin 2011;
 Cao \& Lou 2009, 2010; Lou \& Lian 2011).
Shock wave collapse solutions show the possibility of forming
 black holes within the supernova rebound shock, outside which
 the outer envelopes expand rapidly.
Shock wave solution shown in Figure 8
%{\bf unphysical shock solution?}
 allows inward gas flows while propagating shock energy
 outward due to the self-similar expansion.
It might be possible that rebound shocks in supernovae
 work in this way by some mechanism, avoiding the shock
 difficulties met in unsuccessful explosions of most
 numerical simulations (e.g. Dessart et. al. 2006;
 Buras et. al. 2006; Nordhaus et. al. 2010).
Besides, self-similar dynamic solutions with a central
 expanding void inside the gas sphere may indicate other
 possibilities for the fate of a powerful supernova from
 a massive progenitor;
 if due to some unknown processes certain configurations
 are formed, supernova explosions may leave a much
 rarefied central region, expelling most stellar envelope
 materials outwards (e.g., possibly for SN1987A).
%{\bf Stop here!}

%Equation (\ref{v-diverge}) and (\ref{alpha_vanish}) show that gas being swallowed are greatly accelerated and diluted as approaching the black hole horizon, so all the particles will behave as high energy free particles. For $\gamma>1$ it is seen from equation (\ref{s}) that the sound speed $u_S\sim\sqrt{p/\rho}$ near the black hole tends to zero. Drastic acceleration of atoms and charged particles will produce powerful hard $X-$rays and $\gamma-$rays, leading black hole systems to behave as $X-$ray sources (for black holes of stellar mass), quasars or AGNs (for SMBHs in galaxies). With extremely diluted environments thermal dynamic equilibrium can hardly be achieved, and the produced $X-$rays may have chances to escape with little absorptions. It is known from observations that $X-$ray sources and AGNs have hard radiation spectrums which are power law peaked at short wave length $X-$rays and $\gamma-$rays (Barr et. al. 1980; Osterbrock \& Miller 1975; Koski 1978; Costero \& Osterbrock 1977). For production of hard spectrums the accretion clouds have to be optically thin (Rees et. al. 1982; Shapiro et. al. 1976).

Asymptotic solutions (\ref{v-diverge}) and (\ref{alpha_vanish})
 for $v(x)$ and $\alpha(x)$ show that gas materials being
 swallowed are greatly accelerated and diluted when approaching
 the Schwarzschild black hole, and
 all particles would behave as high-energy free particles.
For polytropic index $\gamma>1$, it is further seen from two
 relations (\ref{s}) and (\ref{m}) that the sound speed
 $u_S=(p/\rho)^{1/2}$ near the black hole event horizon
 tends to vanish.
Drastic infall acceleration of atoms and charged particles would
 give rise to powerful $X-$ray and $\gamma-$ray emissions,
 leading black hole systems to become as astrophysical $X-$ray
 and $\gamma$-ray sources (for stellar mass black holes),
 quasars or AGNs (for SMBHs residing in elliptical galaxies
 and galactic bulges or even HMBHs in clusters of galaxies).
With extremely diluted environments, a thermal dynamic equilibrium
 can hardly be achieved, and the produced $X-$rays and
 $\gamma$-rays may escape from a
 certain inner zone outside the event horizon with relatively weak
 absorptions.
% {\bf but may encounter relatively dense accreting envelope}.
It is known from astrophysical observations for decades that $X-$ray
 sources and AGNs show hard radiation spectra which are power
 laws peaked at short wavelengths of $X-$rays and $\gamma-$rays
 (e.g. Barr et al. 1980; Osterbrock \& Miller 1975; Koski
 1978; Costero \& Osterbrock 1977).
For producing such emissions with hard spectra, the mass
 accretion clouds need to be optically thin (e.g. Rees et
 al. 1982; Shapiro et al. 1976).

\section{Conclusions and Summary}

Gravitational core collapse of a GP gas sphere
 (with EoS $p/\rho^\gamma$ being constant along streamlines)
 towards central Schwarzschild black holes is investigated
 in the self-similar dynamic formalism.
In order to catch key features of the general relativity
 for black holes, we invoke the Paczynski-Wiita
 gravity (Paczynski \& Wiita 1980) instead of the
 Newtonian gravity in our calculations.
% to approximate the general relativity.
A new dimensionless parameter $s$ emerges in the
 self-similar formalism, which corresponds to the square
 of the ratio of the sound speed to the speed of light,
 and the self-similar solutions diverge at the finite
 Schwarzschild radius which expands linearly in time
 with a finite mass accretion rate.
Approaching the event horizon of the central black hole,
 the mass collapse velocity goes to infinity, while the
 gas mass density and pressure vanish.

Various semi-analytical solutions can be constructed
 through numerical shooting integrations by imposing
 proper asymptotic analytic conditions, among which
 the most interesting ones are the EWCSs
%    expansion-wave collapse solutions
 (Shu 1977; Lou \& Shen 2004; Cai \& Shu 2005;
 Wang \& Lou 2008) representing the gravitational
 collapse of static singular GP gas spheres.
In particular, these EWCSs are shown to be closely
 related to the properties of the sonic critical curves.
Cai \& Shu (2005) worked out the EWCS for an isothermal
 gas of $\gamma=1$ in the general relativistic formalism.
%    {\bf Stop here!!} In our conclusions from an
To approximate the general relativity by the the Paczynski-Wiita
 gravity, our GP hydrodynamic model analysis shows
 that for $\gamma<4/3$, only one EWCS exists, while for
 $\gamma>4/3$, there are either two (for small $s<1$) or no
 (for large enough $s<1$) such EWCSs (see Fig. \ref{fig:pcri}
 for more details).
We would expect this conclusion to remain valid within
 the exact formalism of general relativity.
% is fully taken into account.
These EWCSs may offer a sensible description for
 possible gravitational core collapse and accretion of
 nearly static GP gas spheres by the central black hole.
One may also construct various self-similar dynamic
 solutions with non-vanishing radial flow velocities at
 infinity, some of which cross the sonic critical curve
 with or without discontinuity in the first derivatives
 of velocity, while others involve outgoing shocks.
% with jumps in velocity and density at the shock wave front.
For self-similar solutions with high outflow velocities, the
 center of gas sphere may form an expanding void instead of
 a black hole, where both the gas pressure and the mass
 density vanish.

%{\bf Stop here!}

%Spherical ways of accretion may play an considerable role in the quick formation of super-massive black holes (SMBHs) in galaxy nuclei. A black hole with $10^9$ solar mass can be formed within $10^5\sim10^7$ years through spherical accretion, provided there is enough amount of gas or stars. SMBHs are shown to exist in most galaxies (Begelman et al. 1984; Rees 1984; Kormendy \& Richstone 1995), and observations of high red shift quasars (Mortlock et. al. 2011) and AGNs indicate an early formation of SMBHs in $1$ billion years since the Big Bang. Non-uniformly distributed matters and dark matters may results in pre-galaxies and high density peaks, which makes spherical accretion of seeding black holes possible. Matters directly falling into the black hole experiences drastic acceleration and will produce luminous hard $X-$rays or $\gamma-$rays, forming quasars, AGNs or stellar $X-$ray sources.

Quasi-spherical dynamic collapse may play an important role
 of rapidly forming SMBHs in galactic nuclei.
A SMBH of mass $\sim 10^9$ M$_{\odot}$ may be formed within a
 timescale range of $\sim 10^5 -10^7$ years via quasi-spherical
 dynamic mass {\bf collapse}, provided there maintains an enough
 amount of gas materials or stars as the surrounding mass
 reservoirs to sustain such a process.
Accordingly, SMBHs in the mass range of $\sim 10^6-10^9$
 M$_{\odot}$ would be expected to form in a range of shorter
 timescales as compared to the above estimated timescale range.
Of course, such SMBH formation process can be cut short if
 the surrounding mass reservoir cannot be sustained for
 some natural reasons.
Moreover in the cosmological context of large-scale
 structure formation, it is also conceivable to form
 hypermassive black holes (HMBHs) in the mass range of
 $\sim 10^{10}$ M$_{\odot}$ to $\sim 10^{12}$ M$_{\odot}$
 on rare occasions of hypermassive material reservoirs
 (e.g. central regions of rich clusters of galaxies) as
 associated with large-scale mass structures in the
 universe.
Likewise, black holes with mass less than $\sim 10^{6}$
 M$_{\odot}$ may also exist if the quasi-spherical dynamic
 mass {\bf collapse} process is interrupted under various
 situations.
%{\bf Stop here!!}
Observationally, SMBHs seem to lurk in most galactic bulges
 (e.g. Begelman et al. 1984; Rees 1984; Kormendy \& Richstone
 1995), and observations of high red shift (high-$z$) quasars
 (e.g. Mortlock et al. 2011) and AGNs further indicate an
 early formation of SMBHs in about one billion years since
 the Big Bang.
Non-uniformly distributed matters and dark matters may results
 in pre-galaxies, proto-galaxies and high density peaks, which
 may possibly sustain massive reservoirs for quasi-spherical
 dynamic collapses towards seed black holes.
Matters directly falling into the black hole experience drastic
 acceleration and would likely produce luminous hard $X-$ray
 and $\gamma-$ray emissions, forming quasars, AGNs or stellar
 ultra-luminous $X-$ray sources.

%Estimation of the black hole formation during supernova explosions can also be made by self-similar solutions. Collapse of a massive stellar core is believed to lead to black holes (Chandrasekhar 1935; Oppenheimer \& Volkoff 1939; Oppenheimer \& Snyder 1939). Under the extremely relativistic conditions inside a massive supernova with sound speed comparable to light speed, the time scale of the black hole formation is estimated to be $10^{-3}s$, which is $1\sim2$ orders smaller than that of the supernova explosion.
%
%Shock wave solutions also show shock energy is possible to propagate outward in the inflow of gas. We suggest such shocks may appear in the explosion of a massive supernova.
%
%Further, there is possibility that supernova explosion blows all the stellar matter away, leaving a void core inside, as indicated by self-similar solutions with a central void. In this case there may be no compact objects left, as what people suspect for SN1987A.

Estimation of the black hole formation during supernova explosions
 can also be made by self-similar dynamic solutions.
Collapse of a massive
 stellar core is believed to lead to black holes (Chandrasekhar
 1935; Oppenheimer \& Volkoff 1939; Oppenheimer \& Snyder 1939).
Under extremely relativistic conditions inside a massive
 progenitor during supernova with the sound speed less than
 yet comparable to the
 speed of light, the timescale of forming stellar mass black
 holes is estimated to be $\sim 10^{-3}s$, which is $\sim 1$
 to 2 orders of magnitudes shorter than that of the supernova
 explosion.
Self-similar shock solutions also show that the rebound shock
 energy is possible to propagate outward against the inflow
 of gas materials.
We suggest that such shocks may appear in the explosion of
 a massive supernova and leave black holes behind in the
 collapsed core.
By extensive explorations, we note a further physical possibility
 that a supernova explosion blows all the stellar core matter
 away, creating a highly rarefied central void in expansion,
 as implied by self-similar solutions with a central void.
In this case there might be no compact objects left, as has
 been suspected for the absence of central activities in
 SN1987A (e.g. Lou \& Wang 2012).

%\section*{Acknowledgements}
%This research was supported in part by Tsinghua Centre for
% Astrophysics, by the National Natural Science Foundation
% of China grants 10373009, 10533020, 11073014 and J0630317
% at Tsinghua University, by MOST grant 2012CB821800,
%%  射电波段的前沿天体物理课题及FAST早期科学研究 公示内容
% by Tsinghua University Initiative Scientific Research Program,
% and by the Yangtze Endowment, the SRFDP 20050003088 and
% 200800030071, and the Special Endowment for Tsinghua College
% Talent (Tsinghua XueTang) Program from the Ministry of
% Education at Tsinghua University.

\section*{Acknowledgements}
This research was supported in part by the Ministry of
 Science and Technology (MOST) under the State Key
 Development Program for Basic Research grant 2012CB821800,
%   射电波段的前沿天体物理课题及FAST早期科学研究
 by the NSFC
 %National Natural Science Foundation of China
  grants
 10373009, 10533020, 11073014 and J0630317 at Tsinghua University,
 by the Tsinghua University Initiative Scientific Research
 Program, by Tsinghua Centre for Astrophysics, and by the
 Yangtze Endowment, the SRFDP 20050003088, 200800030071 and
 20110002110008, and the Special Endowment for Tsinghua
 College Talent (Tsinghua XueTang) Program from the Ministry
 of Education (MoE) at Tsinghua University.

\appendix

\section{Properties of v(x) Near the Sonic Critical
 Curve and the Zero Mass Line}\label{A}

First, we consider the sonic critical curve with
 $x-v\neq 0$ (i.e., the elliptical-shape loop;
 while the ZML $x-v=0$ and the possible expanding
 void solutions will be considered separately).
%{\bf
The sonic critical curve is determined by
 simultaneously setting both sides of two nonlinear
 ODEs (\ref{ode2}) and (\ref{b1}) to zero.
% Why ODE (\ref{b1})?! }
Therefore, from the LHS of nonlinear ODE (\ref{ode2}),
 we have on the sonic critical curve
\begin{equation}\label{cri1}
\gamma\frac{\beta}{\alpha}=(x-v)^2\ .
\end{equation}
%{\bf
Note that $\beta/\alpha=\alpha^{3\gamma-3}(x-v)^{2\gamma-2}
 x^{4\gamma-4}$; therefore specifically for $1<\gamma<2$,
 we have $\alpha\rightarrow 0$ when approaching the sonic
 critical curve.
% } {\bf What is the point here?!}
 From nonlinear ODE (\ref{b1}), we have
\begin{equation}\label{b1'}
\frac{2}{x}+\frac{1}{\alpha}\frac{\mbox{d}\alpha}{\mbox{d}x}
 =\frac{1}{(x-v)}\frac{\mbox{d}v}{\mbox{d}x}\ .
\end{equation}
Utilizing the L'H\^opital rule and two relations (\ref{cri1})
 and (\ref{b1'}) above, a quadratic equation for the first
 derivative of velocity $\mbox{d}v/\mbox{d}x$ in the vicinity
 of the sonic critical curve is obtained from the vanishing
 two sides of nonlinear ODE (\ref{ode2}), viz.
\[(\gamma+1)\bigg(\frac{\mbox{d}v}{\mbox{d}x}\bigg)^2
 +\bigg[(4\gamma-6)-(4\gamma-4)\frac{(x-v)}{x}\bigg]
 \frac{\mbox{d}v}{\mbox{d}x}\]
 \[\quad +(4\gamma-2)\Big(\frac{x-v}{x}\Big)^2
 -\frac{(2\gamma-2)(4\gamma-2)}{\gamma}\frac{(x-v)}{x}\]
\begin{equation}\label{A3}
\ +\frac{(2\gamma-2)(2\gamma-3)}{\gamma}
 +\frac{m}{(x-sm)^3}\bigg(\frac{x+sm}{x-v}-2\bigg)=0\ ,
\end{equation}
where the reduced enclosed mass $m(x)$ is given
 explicitly by algebraic relation (\ref{m}).
The two roots of $\mbox{d}v/\mbox{d}x$ from equation
 (\ref{A3}) can be properly choosen for distinct
 solutions passing across the sonic critical
 curve smoothly.

We now focus on the ZML $x-v=0$.
 As we always expect a positive enclosed mass
 $m(x)$, the difference $x-v$ should therefore be
 non-negative for a possible physical solution.
Assuming that the first derivative of $v(x)$ near
 the ZML $x-v=0$ approaches a constant value $h$,
 that is
\begin{equation}
\frac{\mbox{d}v}{\mbox{d}x}=h\ ,
\end{equation}
which immediately leads to an approximate expansion
\begin{equation}
x-v=(1-h)(x-x_0)\ ,
\end{equation}
where $x_0$ is the coordinate for our self-similar
 solution to touch the ZML $x-v=0$.
Then by solving nonlinear ODE (\ref{b1'}), we arrive
 at the asymptotic solution for $\alpha(x)$ near the
 ZML in the power-law form of
\begin{equation}
\alpha=\alpha_1(x-x_0)^{h/(1-h)}\ ,
\end{equation}
where $\alpha_1$ is an integration constant. Physically, we would
 require that $m=(x-v)x^2\alpha$ remains
 finite at $x=x_0$ (it should vanish on the ZML, as no
 solution is physically allowed to cross the ZML); thus we simply
 impose the inequality $h<1$.
Substitution of the above asymptotic solution into nonlinear ODE
 (\ref{ode2}) determines the $h$ value.
There are two roots of $h$ parameter. The first root $h_1$ is
\begin{equation}
h_1=\frac{2(1-\gamma)}{\gamma}\le 0\ ,
\end{equation}
with the coefficient $\alpha_1>0$ being an arbitrary constant.
 The second root $h_2$ is given by
\begin{equation}
h_2=\frac{2(2-\gamma)}{(\gamma+1)}\ge 0\ ,
\end{equation}
where the constant coefficient $\alpha_1$
 satisfies the condition
\begin{equation}
h_2\big[1-h_2\big]^{4-2\gamma}=\big[\gamma
 h_2+2\gamma-2\big]
 \alpha_1^{3\gamma-3}x_0^{4\gamma-4}\ .
\end{equation}

\section{Derivation of equation
   for the sonic critical curve}

The sonic critical curve of two coupled nonlinear ODEs
 (\ref{ode1}) and (\ref{ode2}) can be readily derived
 by setting the both sides of these two nonlinear ODEs
 to vanish, namely
\begin{equation}\label{c1}
(x-v)^2-\gamma\frac{\beta}{\alpha}=0\ ,
\end{equation}
\begin{equation}\label{c2}
\frac{(x-v)x^2\alpha}{[x-s(x-v)x^2\alpha]^2}
 +\frac{(2\gamma-2)}
 {(x-v)}\frac{\beta}{\alpha}-\frac{2(x-v)^2}{x}=0\ ,
\end{equation}
\begin{equation}\label{c3}
\frac{(x-v)x^2\alpha}{[x-s(x-v)x^2\alpha]^2}
 +\bigg[\frac{(2\gamma-2)}{(x-v)}
 -\frac{2\gamma}{x}\bigg]\frac{\beta}{\alpha}=0\ ,
\end{equation}
where the reduced gas pressure
 $\beta=\alpha^{3\gamma-2}(x-v)^{2\gamma-2}x^{4\gamma-4}$
 is defined by GP EoS (\ref{beta}) according
 to the specific entropy conservation along streamlines.
In reference to equation (\ref{c1}), only one of the two
 algebraic equations (\ref{c2}) and (\ref{c3}) is independent;
 this would guarantee the consistency of our model analysis.
Eliminating the reduced mass density $\alpha$ by proper
 substitutions, we reach the following algebraic equation
 for the sonic critical curve
\begin{equation}
\frac{x-v}{x}-\frac{\gamma-1}{\gamma}=
 \frac{\gamma^{\frac{1}{3-3\gamma}}
 (x-v)^{\frac{\gamma+1}{3\gamma-3}-1}x^{2/3}}
 {2\big[x-s\gamma^{\frac{1}{3-3\gamma}}
 (x-v)^{\frac{\gamma+1}{3\gamma-3}}x^{2/3}\big]^2}
\end{equation}
as stated in the main text.

%{\bf Newly added at Narita according to Lian Biao April 18 email}

%\section{Figure 1: Number of Roots}
%Let us define function $f(\alpha_0)=2\alpha_0^{3\gamma-4}(1-s\alpha_0)^2$, so our task is to solve equation $f(\alpha_0)=1$. For physical purposes we must demand $1-s\alpha_0>0$, thus physical roots must locate on the interval $(0,1/s)$.

%It is easily seen for $\gamma<4/3$, both $\alpha_0^{3\gamma-4}$ and $(1-s\alpha_0)^2$ are monotonously decreasing, so $f(\alpha_0)$ keeps decreasing from infinity to zero on the interval $(0,1/s)$, and therefore only one root exists.

%For $\gamma>4/3$, $\alpha_0^{3\gamma-4}$ is increasing. $f(\alpha_0)$'s derivative is
%\begin{equation}
%\frac{\mbox{d}f}{\mbox{d}\alpha_0}(\alpha_0)=
%2\alpha_0^{3\gamma-4}(1-s\alpha_0)^2\Big(\frac{3\gamma-4}{\alpha_0}-\frac{2}{1-s\alpha_0}\Big)\ ,
%\end{equation}
%so $f(\alpha_0)$ first increase then decrease, taking zero value at both $\alpha_0=0$ and $\alpha_0=1/s$. It then reaches its maximum at $\alpha_{0M}=(3\gamma-4)/[2+(3\gamma-4)s]$, where its value is
%\begin{equation}
%f_{max}=\frac{8(3\gamma-4)^{3\gamma-4}}{[2+(3\gamma-4)s]^{3\gamma-2}}\ .
%\end{equation}
%For $f_{max}>1$, there are two roots. For $f_{max}<1$, there will be no roots. The numerical results are shown in Figure 1. Specifically for $s<1-1/\sqrt{2}=0.2929$, $f_{max}$ is always greater than $1$, which ensures the existence of two roots.

\section{Number of Roots in Three Distinct Regimes of Figure 1}

By defining a function $f(\alpha_0)\equiv
  2\alpha_0^{3\gamma-4}(1-s\alpha_0)^2$ in
  reference to equation (\ref{alpha0'}), our task
  is to actually solve equation $f(\alpha_0)=1$
  for possible roots of $\alpha_0$.
For astrophysical applications, we should demand
  inequality $1-s\alpha_0>0$, and thus physical roots
  of $\alpha_0$ fall within the open interval $(0,\ 1/s)$.

It is easily seen for $\gamma<4/3$, both $\alpha_0^{3\gamma-4}$
 and $(1-s\alpha_0)^2$ decrease monotonically as $\alpha_0$
 increases; therefore, $f(\alpha_0)$ keeps decreasing from
 infinity to zero within the open interval $(0,\ 1/s)$ and
 only one $\alpha_0$ root exists.

For $\gamma>4/3$, $\alpha_0^{3\gamma-4}$
 increases as $\alpha_0$ increases.
The first derivative of $f(\alpha_0)$ with
 respect to $\alpha_0$ is simply
\begin{equation}
\frac{\mbox{d}f}{\mbox{d}\alpha_0}(\alpha_0)=
 2\alpha_0^{3\gamma-4}(1-s\alpha_0)^2\bigg[\frac{(3\gamma-4)}
 {\alpha_0}-\frac{2s}{(1-s\alpha_0)}\bigg]\ ,
\end{equation}
indicating that $f(\alpha_0)$ first increases and then
 decreases with increasing $\alpha_0$, taking zero
 values at both $\alpha_0=0$ and $\alpha_0=1/s$.
Function $f(\alpha_0)$ then reaches its maximum value
 $f_{\rm max}$ at $\alpha_{0M}=(3\gamma-4)/
 [(3\gamma-2)s]$ with $f_{\rm max}$ given by
\begin{equation}
f_{\rm max}=\frac{8[(3\gamma-4)/s]^{3\gamma-4}}
 {(3\gamma-2)^{3\gamma-2}}\ .
\end{equation}
For $f_{\rm max}>1$, there exist two roots for $\alpha_0$.
For $f_{\rm max}<1$, there will be no roots for $\alpha_0$.
 For $f_{\rm max}=1$, the two roots of $\alpha_0$ become
 degenerate.
The pertinent numerical results are displayed in Figure 1.
Specifically for $s<1-1/\sqrt{2}=0.2929$, the maximum
 value $f_{\rm max}$ is always greater than $1$, which
 ensures the existence of two roots for $\alpha_0$.
For other pairs of $\gamma$ and $s$ in the type B
 regime of Figure \ref{fig:pcri}, there are also
 two roots for $\alpha_0$.
One needs to check whether the two $\alpha_0$ roots
 are less than $1/s$ to be physically useful.

%\section{Derivation of equation of critical line}
%The critical line of ODEs (\ref{ode1}) and (\ref{ode2}) can be derived out by setting both sides of the two ODEs to zero, namely
%\begin{equation}\label{c1}
%(x-v)^2-\gamma\frac{\beta}{\alpha}=0\ ,
%\end{equation}
%\begin{equation}\label{c2}
%\frac{(x-v)x^2\alpha}{[x-s(x-v)x^2\alpha]^2}+\frac{2\gamma-2}{x-v}\frac{\beta}{\alpha}-\frac{2(x-v)^2}{x}=0\ ,
%\end{equation}
%\begin{equation}\label{c3}
%\frac{(x-v)x^2\alpha}{[x-s(x-v)x^2\alpha]^2}+\Big(\frac{2\gamma-2}{x-v}-\frac{2\gamma}{x}\Big)\frac{\beta}{\alpha}=0\ ,
%\end{equation}
%where $\beta=\alpha^{3\gamma-2}(x-v)^{2\gamma-2}x^{4\gamma-4}$, as defined in equation (\ref{beta}). Note that only two of the above three equations are independent. Eliminate the variable $\alpha$ by substitution, we reached the equation of critical line
%\begin{equation}
%\frac{(x-v)^2}{x}-\frac{\gamma-1}{\gamma}(x-v)=\frac{\gamma^{\frac{1}{3-3\gamma}}(x-v)^{\frac{\gamma+1}{3\gamma-3}}x^{2/3}} {2[x-s\gamma^{\frac{1}{3-3\gamma}}(x-v)^{\frac{\gamma+1}{3\gamma-3}}x^{2/3}]^2}\ .
%\end{equation}

\bsp

\label{lastpage}

\end{document}